\documentclass[12pt]{article}
\usepackage{epsfig}\parskip 5pt plus 1pt
\usepackage{bbm}
\usepackage{amsmath}
\usepackage{amssymb}
\usepackage{amsfonts}
\usepackage{mathrsfs}
\usepackage{graphicx}
\usepackage{axodraw}
\usepackage{color}
\newcommand{\nn}{\nonumber}

\newcommand{\dv}{\partial\hspace{-7pt}\slash}

\newcommand{\D}{D\hspace{-8pt}\slash}

\newcommand{\be}{\begin{equation}}
\newcommand{\ee}{\end{equation}}
\newcommand{\bea}{\begin{eqnarray}}
\newcommand{\eea}{\end{eqnarray}}

\begin{document}
\begin{titlepage}
\vspace*{-1cm}
\flushright{FTUAM-07-12\\ IFT-UAM/CSIC-07-41\\ LPT-Orsay 07-34\\ULB-TH/07-27}
\vskip 2.0cm
\begin{center}
{\Large\bf Low energy effects of neutrino masses}
\end{center}
\vskip 1.0  cm
\begin{center}
{\large A. Abada}$\,^a$~\footnote{asmaa.abada@th.u-psud.fr},
{\large C.Biggio}$\,^b$~\footnote{carla.biggio@uam.es},
{\large F. Bonnet}$\,^a$~\footnote{florian.bonnet@th.u-psud.fr}\\
\vskip .1cm
{\large M.B. Gavela}$\,^b$~\footnote{belen.gavela@uam.es} and 
{\large T. Hambye}$\,^{b,c}$~\footnote{thomas.hambye@uam.es}\\
\vskip .2cm
$^a\,$ Laboratoire de Physique Th\'eorique  UMR 8627,\\
Universit\'e de Paris-Sud 11, Bat. 210, 91405 Orsay Cedex, France
\vskip .1cm
$^b\,$ Departamento de F\'\i sica Te\'orica and Instituto de F\'\i sica Te\'orica IFT-UAM/CSIC,
\\
Universidad Aut\'onoma de Madrid, 28049 Cantoblanco, Madrid, Spain
\vskip .1cm
$^c\,$ Service de Physique Th\'eorique,\\
Universit\'e Libre de Bruxelles, 1050 Brussels, Belgium
\end{center}
\vskip 0.5cm
\begin{abstract}

While all models of Majorana neutrino masses lead to the same dimension five effective operator, which does not conserve lepton number, the dimension six operators induced at low energies conserve lepton number and differ depending on the high energy model of new physics. We derive the low-energy dimension six operators which are characteristic of generic Seesaw models, in which neutrino masses result from  the exchange of heavy fields which may be either  fermionic singlets, fermionic triplets or scalar triplets. The resulting  operators  may lead to  effects observable in the near future, if the coefficients of the dimension five and six operators are decoupled  along a certain pattern, which turns out to be common to all models. The phenomenological consequences are explored as well, including their contributions to $\mu \rightarrow e \gamma$ and new bounds on the Yukawa couplings for each model.
\noindent
\end{abstract}
\end{titlepage}
\setcounter{footnote}{0}
\vskip2truecm
\newpage
\tableofcontents
\newpage
\section{Introduction}
\label{intro}
The experimental observation of non-zero neutrino masses and mixings constitutes  evidence for physics beyond the Standard Model (SM) and points to the existence of a new, yet unknown, physics scale. It has been already a few years since the breaking of such exciting news and nevertheless little -if anything- is known about the underlying physics. The difficulty lies in both the fact that neutrinos are very weakly interacting particles
 and, more important, in the tiny value of their masses - orders of magnitude lighter than any other fermion masses -  pointing to very suppressed effects. The absence of exotic experimental signals other than neutrino masses, as well as the theoretical criteria of naturalness, point to values of the new physics scale, $M$,  larger than  the electroweak scale.

  It is worth recalling that the evidence for neutrino masses comes from neutrino oscillations, which detect the interference between the different paths taken by different neutrinos when traveling a long distance. The paths differ because the masses differ and what has been  measured is the relative phase shift induced, which is only sizable after extremely long distances. In other words, detection has been possible because neutrino masses affect neutrino propagation. Other possible low-energy effects of the underlying theory, i.e. exotic couplings, are typically zero-distance effects 
   which cannot benefit from such an enhancement. 
 Its suppression is only easily overcome at very high energies, with the particle momenta equal or larger than the scale $M$, as for instance 
 in leptogenesis scenarios, where  the high energies of the early universe allow the heavy fields  at the origin of neutrino masses to roam freely.
 
  
   To see what could be the nature and magnitude of the low energy effects associated to neutrino masses it is convenient to rephrase
  the above in terms of a generic effective low-energy theory. 
  Effective theories allow rather model-independent analysis based on the fundamental symmetries, while only the coefficient of the effective operators are  model-dependent.
 The impact at low energies of the heavy fields present  in  the putative high-energy theory  
 can be parametrized, without loss of generality, by an effective Lagrangian including:
 \begin{itemize}
 \item Corrections to the parameters of the SM Lagrangian.
 \item The addition to the SM Lagrangian of a tower  of {\it non-renormalizable} higher-dimension 
 operators, invariant under the SM gauge group. The latter are made out of the SM fields active at low energies
  and their coefficients weighted by inverse powers of the high scale $M$, 
 \be\label{leff}
{\cal L}_{\rm eff} = {\cal L}_{\rm SM} + \delta{\cal L}^{d=5} + \delta{\cal
L}^{d=6} + \cdots
\ee
 \end{itemize}
 The only possible dimension $5$ ($d=5$) operator   is the famous Weinberg operator~\cite{Weinberg:1979sa},
 \begin{equation}
 \delta{\cal L}^{d=5} = \frac{1}{2}\, c_{\alpha \beta}^{d=5} \,
\left( \overline{\ell_L^c}_{\alpha} \tilde \phi^* \right) \left(
\tilde \phi^\dagger \, {\ell_L}_{ \beta} \right) + {\rm h.c}.\, ,
\label{d=5}
\end{equation}
where $\ell_L$ stands for the lepton weak doublets\footnote{The
charge-conjugate spinor is denoted ${\psi}^c \equiv C
\overline{\psi}^T$, where $T$ denotes transposition and $C$ charge
conjugation.}, greek letters denote flavour indices and $\tilde \phi$
is related to the standard Higgs doublet $\phi\equiv (\phi^+, \phi^0)$
by $\tilde \phi = i \tau_{2} \phi^*$.  Finally, $c_{\alpha
\beta}^{d=5}$ is a coefficient matrix of inverse mass dimension, i.e.
${\cal{O}}(1/M)\,$.  This operator is not invariant under the $B-L$
symmetry, with $B$ and $L$ denoting respectively baryon and lepton
number, which is an accidental symmetry of the SM. Upon electroweak
symmetry breaking, $<\phi^0> =v/\sqrt{2}$, $v= 246$ GeV, this term
results in Majorana neutrino masses.  
Such a $d=5$ operator is
characteristic of all theories with Majorana neutrino masses, such as
for instance the minimal (type I) Seesaw
model~\cite{Seesaw}. Therefore, the knowledge of $c_{\alpha
\beta}^{d=5}$ doesn't allow to discriminate between these models. It
is very suggestive that the lowest-order effect of high-energy beyond
the Standard Model physics may be neutrino masses. There is no hope to
see any other low energy effects, e.g.~zero distance effects,
associated to this operator. These effects are necessarily tiny since
neutrino masses - which fix the $c_{\alpha \beta}^{d=5}$ coefficients
- are tiny\footnote{Notice that neutrino masses have been detected in
neutrino oscillation experiments, which in fact measure differences
between the {\it square} of neutrino masses. That is, if the neutrinos
are Majorana particles, the experiments have already measured an
effect suppressed as $(c_{\alpha \beta}^{d=5})^2\sim1/M^2$ instead of
$1/M$ and thus quantitatively alike to that from generic dimension six
operators.}.
  
 The case of the  dimension six ($d=6$) $SU(3)\times SU(2) \times U(1)$ invariant operators is different, though.
There is a plethora  of such operators~\cite{Buchmuller:1985jz}. 
Different classes of models result in different $d=6$ operators. 
 Their  identification and eventually their experimental selection is then a very important tool to  discriminate the origin of neutrino masses. 
  An important property of these operators  is that their coefficients are not necessarily as suppressed as that for the $d=5$ operator and, therefore, may lead to observables low-energy effects. The point is that {\it all $d=6$ operators preserve $B-L$, in contrast with the unique $d=5$ operator above. This suggests that, from the point of view of symmetries, it may be natural to consider large coefficients for the $d=6$ operators resulting from the new physics, while having small coefficients for the $B-L$ odd operator}.
  Such a possibility would require to decouple the coefficients of the $d=6$ operators from that of the $d=5$ operator  responsible for neutrino masses.

\begin{figure}[tbh]
\centering
\begin{picture}(400,100)
\DashLine(10,80)(40,50){4}
\ArrowLine(10,20)(40,50)
\ArrowLine(40,50)(60,50)
\ArrowLine(80,50)(60,50)
\DashLine(80,50)(110,80){4}
\ArrowLine(110,20)(80,50)
\Text(60,60)[]{$N_R$}
\Text(120,20)[]{$\ell$}
\Text(120,80)[]{$\phi$}
\Text(95,50)[]{$Y_N$}
\Text(25,50)[]{$Y_N^\dagger$}
\Text(0,80)[]{$\phi$}
\Text(0,20)[]{$\ell$}
\DashLine(170,100)(200,70){4}
\ArrowLine(170,0)(200,30)
\DashLine(200,70)(200,30){4}
\DashLine(200,70)(230,100){4}
\ArrowLine(230,0)(200,30)
\Text(165,100)[]{$\phi$}
\Text(165,0)[]{$\ell$}
\Text(240,100)[]{$\phi$}
\Text(240,0)[]{$\ell$}
\Text(210,50)[]{$\Delta$}
\Text(200,85)[]{$\mu_\Delta$}
\Text(200,15)[]{$Y_\Delta$}
\DashLine(290,80)(320,50){4}
\ArrowLine(290,20)(320,50)
\ArrowLine(320,50)(340,50)
\ArrowLine(360,50)(340,50)
\DashLine(360,50)(390,80){4}
\ArrowLine(390,20)(360,50)
\Text(340,60)[]{$\Sigma_R$}
\Text(400,20)[]{$\ell$}
\Text(400,80)[]{$\phi$}
\Text(375,50)[]{$Y_\Sigma$}
\Text(305,50)[]{$Y_\Sigma^\dagger$}
\Text(280,80)[]{$\phi$}
\Text(280,20)[]{$\ell$}
\end{picture}
\caption{The three generic realizations of the Seesaw mechanism, depending on the nature of the 
heavy fields exchanged: SM singlet fermions (type I Seesaw) on the left, SM triplet scalars (type II Seesaw) and SM triplet fermions (type III Seesaw) on the right.}
\label{trois-types}
\end{figure}
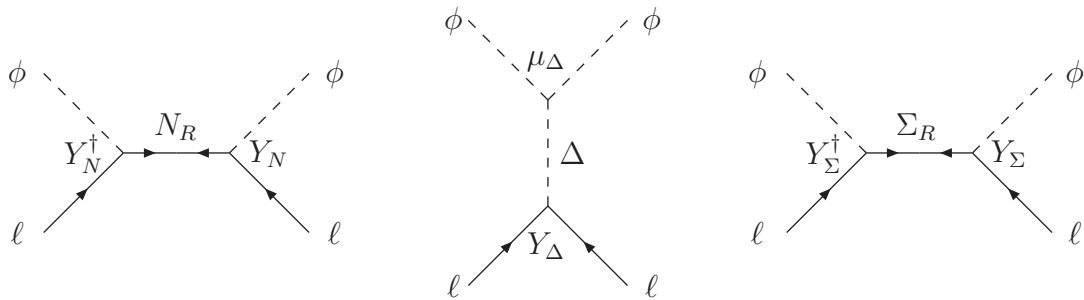

    The first purpose of this work is to identify the effective $d=6$ operators which are characteristic of Seesaw models (Section 2). 
 In the latter, the tiny neutrino masses naturally result  from the tree-level exchange of heavy particles, which may be either fermions or bosons. 
 The exchange of heavy SM singlet fermions is the essence of  the minimal Seesaw model (type I) and its generalizations. Analogously, the exchange of heavy $SU(2)_L$ scalar triplets is another possibility which has been widely explored, as in the type II Seesaw model and its generalizations~\cite{type-II}. $SU(2)_L$ fermionic triplets may also mediate light neutrino masses (type III Seesaw)~\cite{type-III,Ma:2002pf,Hambye:2003rt,bajc,fileviez}. 
 Most beyond the SM  theories with Majorana neutrino masses typically incorporate one of these mechanisms or combinations of them: the lessons learnt from their study should  be of extensive relevance.
 We will thus discuss the effective low-energy Lagrangians for the three generic cases: heavy fermion singlets, heavy  scalar triplets and heavy  fermionic triplets,  illustrated in Fig. \ref{trois-types}.

  Next, in a second stage (Section 3) we consider the possibility that the 
  $d=6$ operators are not as suppressed as the $d=5$ operator, so that observable
  low-energy effects may be expected. Since these operators are suppressed  by $1/M^2$, this requires a value of $M$ not far beyond the electroweak scale.
  We consider this possibility, which is not excluded at all and may even be supported by hierarchy arguments.
  It will then be shown that in order to have observable low energy effects, it is  necessary and possible to decouple and suppress the coefficient of the $d=5$ operator relative to the $d=6$ operator coefficients, in a way which accommodates tiny neutrino masses while allowing large Yukawa couplings.
   It will be shown that such decoupling  requires a common and rather model-independent pattern, which we identify.
   
   In a third stage (Section 4) and independently of how large is the
scale $M$, we analyze the long list of phenomenological signals which
may arise in each of the three models considered, such as signals
associated to non-unitarity or other effects in different observables:
neutrino oscillations, lepton and gauge-boson decays.  From present
data, limits will be set in all models on the coefficients of the
$d=6$ operators.  From them, we derive systematic tables of bounds on
the Yukawa couplings in each of the Seesaw models.  Expectations for
the sensitivity of future experiments will be explored, including the
contributions to $l_i\rightarrow l_j \gamma$. We show that, in case
the decoupling pattern mentioned above occurs, the limits can be
saturated if $M$ is still larger than but close to the electroweak
scale. The possibilities for direct or indirect discovery of the
origin of neutrino masses at the LHC or ILC will be (briefly)
discussed.
   
    An important phenomenon at the origin of many of the potential low
 energy effects is non-unitarity of the leptonic mixing
 matrix. Special emphasis will be set on analyzing whether Seesaw
 models induce at low energies a non-unitary leptonic mixing
 matrix. It is expected in all generality~\cite{uni} that the
 tree-level exchange of heavy fermions (scalars) will (not) induce it.
 Indeed, only leptons can mix with other fermions leading to (unitary)
 mixing matrices of dimension larger than $3$, while the submatrix for
 the light fields needs not be unitary. In a more technical view, the
 exchange of heavy fermions among light leptons can be understood from
 the expansion of the heavy field propagator in powers of $1/M$, \be
 \frac{1}{\D -M}\,\sim\, -\frac{1}{M}\,+\, \frac{1}{M}\D\,
 \frac{1}{M}+...  \ee The first term in this expansion is a scalar
 operator, which flips chirality, generating for instance a light
 neutrino mass term. The second term, instead, preserves chirality and
 induces a correction to the kinetic term for the light fields. The
 recovery of canonically normalized kinetic energies for the latter
 requires in general a flavour-dependent rescaling, which is a
 non-unitary transformation, surfacing as non-unitary mixing matrices
 in the leptonic weak currents~\cite{uni}.  Non-unitarity of the
 leptonic mixing matrix is therefore a basic property of models where
 masses are induced by heavy fermions.  In contrast, in
 scalar-mediated mechanisms, all terms in the scalar propagator change
 chirality and thus cannot induce non-unitary mixing at
 tree-level. The minimal (type I) Seesaw model has been previously
 shown~\cite{BGJ} to induce a non-unitary leptonic mixing matrix. In
 this work we will explicitly analyze the issue for the other types of
 Seesaw models.

\newpage
\section{The basic Seesaw scenarios}
\label{largeM}
Let us analyze separately the three different minimal models which result from adding either fermionic singlets or scalar triplets or fermionic triplets to the minimal SM field content. It is expected that the lessons obtained from the analysis of the three basic models will hold as well for their possible generalizations, extensions or embeddings in larger theories.
\subsection{Fermionic singlets: Type I Seesaw}
\label{typeI}
As this case has been previously studied~\cite{BGJ}, only the main results are resumed here for completion.
The minimal Seesaw Lagrangian is the most general renormalizable Lagrangian which
can be written for the SM gauge group adding only right-handed neutrinos
to the SM fermion content of the theory.
The leptonic Lagrangian of the Seesaw model 
is given by
\be
\label{SS_lagr}
{\cal L}_{\rm leptons}= {\cal L}^{\rm KE}_{\rm leptons}
+{\cal L}^{ \rm SB}_{\rm leptons},
\ee
where 
\be
{\cal L}^{\rm KE}_{\rm leptons} = i \, 
\overline{\ell_L}
 \,
\D \,
 \ell_L
  + i \, \overline{e_R} \, \D \, e_R
+ i \, \overline{N_R} \, \dv \, N_R  \\
\ee
contains the kinetic energy and gauge interaction terms 
of the left-handed lepton doublets 
$\ell_L$, 
the right-handed charged leptons $e_R$, 
the right-handed neutrinos $N_R$ and
\be
{\cal L}^{ \rm SB}_{\rm leptons} =
-  \overline{\ell_L} \,{\phi} \, {Y_e}  \, e_R
- \overline{\ell_L} \,{\widetilde\phi} \, {Y_N^\dagger}  \, N_R
-\frac{1}{2}\,\overline{N_R} \, {M_N} \,{{N_R}^c}
+\text{h.c.}
\ee
contains the Yukawa interactions with coupling $Y_N$ and the Majorana mass term of 
the gauge-singlet right-handed neutrinos, corresponding to the new physics scale(s) $M_N$. Flavour indices are implicit in these expressions and we will work in a basis in which $M_N$ is a diagonal complex matrix.
\subsubsection{Dimension 5 operator}
In the flavour basis, the resulting $d=5$ operator coefficients
are given in terms of the parameters of the
high-energy theory as (see Fig.~1)
\bea\label{cd5}
c^{d=5} =  Y_N^T \,\frac{1}{ M_N} \, Y_N \,.
\eea  
Upon electroweak symmetry breaking, 
it leads to a Majorana mass matrix
for the light neutrinos of the form 
 \be\label{mnuI}
m_\nu\equiv -{v^2 \over {2}} \, c^{d=5}\,= -{ 1\over {2}} \,Y_N^T \frac{v^2} { M_N} \, Y_N \,.
\ee 
For values of the Yukawa couplings $Y_N$ of order unity, 
the tiny experimental values of neutrino masses require a scale $M_N$ suggestively close to the Grand Unification scale.
\subsubsection{Dimension 6 operator}
In Ref.~\cite{BGJ}, the $d=6$ low-energy effective theory, $\delta {\cal L}^{d=6}$, was determined to consist at the tree level of  the unique operator
\be\label{d6}
\delta{\cal L}^{d=6} = c^{d=6}_{\alpha \beta} \, \left( \overline{\ell_{L\alpha}} \tilde \phi
\right) i \dv \left( \tilde \phi^\dagger \ell_{L \beta} \right),
\ee 
where the $d=6$ operator coefficients are given in terms of the parameters
of the high-energy Seesaw theory by
\be\label{cd6}
c^{d=6} = Y_N^\dagger \, \frac{1}{M_N^\dagger}\frac{1}{M_N} \, Y_N \,,
\ee  
which is of the same order in Yukawa couplings than its $d=5$ counterpart, Eq.~(\ref{cd5}), while quadratically suppressed in $1/M_N$.
When the Higgs doublet acquires a vacuum
expectation value, this $d=6$ operator leads to corrections to the $d=4$ kinetic energy terms for
the left-handed Majorana neutrinos,
which result in a non-unitary low-energy leptonic mixing matrix~\cite{uni}. 
Indeed, the neutrino Lagrangian for the effective theory, including only
$d \le 6$ operators and disregarding couplings to the physical Higgs particle,  is given by
\be
\label{d6_lagr}
{\cal L}_{\rm neutrino}^{d\le6}= i\,\overline\nu_{L \alpha}\,\dv \,
\left(\delta_{\alpha \beta}+ \epsilon^N_{\alpha \beta} \right) \, \nu_{L \beta}
-\frac{1}{2}
\overline{{\nu_L}^c}_\alpha\,m_{\nu\,\alpha\beta}\,{\nu_L}_\beta
-\frac{1}{2}\overline{{\nu}_L}_\alpha\,m^*_{\nu\,\alpha\beta}\,
{{\nu_L}_\beta}^c~,
\ee
where 
\be
\label{epsilonN}
\epsilon^N \equiv {v^2 \over {2}} \, c^{d=6}
\ee
is the contribution of the
$d=6$ operator coefficient to the left-handed neutrino kinetic energy, which is 
non-diagonal in flavor  space.  Let us then go  to a basis in which the neutrino field is rescaled, so that
the neutrino kinetic energy is canonically normalized: at order ${\cal O}(1/M^2)$, the transformation 
\begin{eqnarray}
{\nu_{L}}_{\alpha}&\rightarrow&{\nu'_{L}}_{\alpha}\equiv \left(\delta_{\alpha\beta}+\epsilon^{N}_{\alpha\beta}\right)^{\frac{1}{2}}{\nu_{L}}_{\beta}\,
\end{eqnarray}
results in a Lagrangian in the flavour basis which, at this order, takes the form (primes will be omitted in the following),
\begin{eqnarray}
\label{LagShift}
\mathcal{L}^{d\leq6}_{\textrm{leptons}} = i\overline{\nu_{L}}_{\alpha}\partial\!\!\!/ {\nu_{L}}_{\alpha}+i\overline{l_{L}}_{\alpha}\partial\!\!\!/{l_{L}}_{\alpha} 
-\frac{1}{2}
\left[\overline{{\nu_L}^c}_\alpha\,m_{\nu\,\alpha\beta}\,{\nu_L}_\beta
+ \text{h.c.}\right] -\overline{l_\alpha}m_{l\, \alpha \beta} l_\beta
+ \mathcal{L}_{CC} + \mathcal{L}_{NC}+\mathcal{L}_{em}\,,
\end{eqnarray}
where $m_l$ is the charged lepton mass matrix and 
\begin{eqnarray}
\label{JI}
\mathcal{L}_{CC} &=&\frac{g}{\sqrt{2}}\overline{l_{L}}_{\alpha}{W\!\!\!\!\!/}\;^{-}\left(\delta_{\alpha\beta}-\frac{1}{2}\epsilon^{N}_{\alpha\beta}\right){\nu_{L}}_{\beta}+ \text{h.c.}\,,\\
\mathcal{L}_{NC}&=&\frac{g}{cos\theta_{W}}\left\{\frac{1}{2}\left[\overline{\nu_{L}}_{\alpha}\gamma_{\mu}\left(\delta_{\alpha\beta}-\epsilon^{N}_{\alpha\beta}\right){\nu_{L}}_{\beta}-\overline{l_{L}}_{\alpha}\gamma_{\mu}{l_{L}}_{\alpha}\right]-sin^{2}\theta_{W}J^{em}_{\mu}\right\}Z^{\mu}\nonumber \,,\\
\mathcal{L}_{em}&=&eJ^{em}_{\mu}A^{\mu}\nonumber\,,
\end{eqnarray}
with $J^{em}_{\mu}=-\overline{l}\gamma_\mu l$ denoting the electromagnetic current. 
We can now rotate to the basis in which 
the mass matrices are diagonal,
\bea
{\cal L}_{\rm leptons}^{d\le 6}=
\frac{1}{2}\overline {\nu_i} \left( i\dv- m^{diag}_{\nu\,i }\right) {\nu_i}+\frac{1}{2}\overline {l_i} \left( i\dv- m^{diag}_{l\,i }\right) {l_i}+ \mathcal{L}_{CC} + \mathcal{L}_{NC}+\mathcal{L}_{em}\,.
\eea
 Now, because of the flavour-dependent field rescalings  involved, the usual 
 $U_{PMNS}$ matrix appearing in the charged-current coupling is replaced by a non-unitary matrix $N$, 
\bea
\label{NN}
N\equiv \Omega\, \left(1-\frac{\epsilon^{N}}{2}\right)\,U^\nu\,,
\eea
where $U^\nu$ diagonalizes the neutrino mass matrix
 and $\Omega\equiv{\rm diag}(e^{i\omega_1},e^{i\omega_2},e^{i\omega_3})$  reabsorbs three unphysical phases in the definition of the charged lepton fields, as usual. Details of the procedure can be found in Appendix A. Notice that,
 as $U^\nu$ does not depend on $c^{d=6}$ at ${\cal O}(1/M^2)$, in a flavour basis in which $\Omega$ is the identity matrix, $N$ would read 
 \begin{equation}
 \label{NNbasis}
 N= \left (1- {\epsilon^N \over 2}\right)U_{PMNS}\, 
\end{equation}
and consequently $N N^\dagger=(1-\epsilon^N)$, $N^\dagger N= U_{PMNS}^\dagger(1-\epsilon^N)U_{PMNS}$, within the ${\cal O} (1/M_N^2)$ considered in this work.
 
 Whatever the flavour basis, in the mass basis  the weak currents read now
\bea
\label{JCC_d6}
J_\mu^{-\,CC}  
&\equiv& \overline {e_L}_\alpha \, \gamma_\mu \,
N_{\alpha i} \, \nu_i,\\
\label{JNC_d6}
J_\mu^{NC}
&\equiv& {1 \over 2} \overline \nu_i \,\gamma_\mu
(N^\dagger \,N)_{i j}
\,\nu_j ,
\eea
where $\sum_\alpha {N^\dagger}_{i \alpha}\,N_{\alpha j} \ne
\delta_{ij}$ appears in the neutral current since $N$ is not
unitary, while 
 the neutral current for charged leptons is the standard one. Accordingly, the Fermi constant measured in experiments, $G_F$, cannot be identified anymore with the SM tree level combination $G_F^{SM} = \sqrt{2} g^2 / (8 M_W^2)=\frac{1}{\sqrt{2} v^2}$, due to non-unitarity. For instance, the Fermi constant $G_F$ extracted from the decay $\mu \rightarrow \nu_\mu e \bar{\nu}_e$ is related to $G_F^{SM}$ by~\cite{uni}
\be
\label{Eq:GFM}
G_F = 
G_F^{SM}
 \sqrt{(NN^\dagger)_{ee} (NN^\dagger)_{\mu\mu} }\, .
\ee
The rest of the parameters of the Lagrangian coincide with those in the standard treatment.
 It is remarkable that putative departures from unitarity of the
 leptonic mixing matrix can be now directly related to the $d=6$
 operator coefficients and thus to combinations of the high-energy
 parameters\footnote{In the flavour basis above mentioned, in which
 $N$ is given by Eq.~(\ref{NNbasis}) and the matrix $\Omega$ is the
 identity, the absolute-value bars can be dropped: $(NN^\dagger
 -1)_{\alpha\beta} \,=\,\frac{v^2}{2}\,
 c^{d=6}_{\alpha\beta}\,=\,\frac{v^2}{2}\,(Y_N^\dagger
 \frac{1}{M_N^\dagger}\frac{1}{ M_N}{Y_N})_{\alpha\beta}\,.$},
\begin{equation}
\label{Nc6-N}
|NN^\dagger  -1|_{\alpha\beta} \,=\,\frac{v^2}{2}\,|c^{d=6}|_{\alpha\beta}\,=\,\frac{v^2}{2}\,|Y_N^\dagger\frac{1}{M_N^\dagger}\frac{1}{ M_N}Y_N|_{\alpha\beta}\,.
\end{equation} 
In Sect.~\ref{typeIpheno} the present numerical constraints on $|c^{d=6}|$ will be explored.
 For Yukawa couplings $Y_N\sim \cal{O}$$(1)$, the coefficients of the $d=6$ operator are basically the square of those for the $d=5$ operator, as Eqs.~(\ref{cd5}) and (\ref{cd6}) show. The smallness of neutrino masses then requires e.g.~${M_N}\gg v$,
 which precludes the observation of exotic effects in present and planned facilities for the minimal model discussed in this Section. There are, however, situations in which $Y_N\sim{\cal O}(1)$ can be accommodated together  with 
  $M_N\sim {\mathcal O} (TeV)$ and without fine-tunings, leading to observable effects in the near future, as it will be discussed in Section~\ref{lowscale}.
  
In Ref.~\cite{BGJ},  it was shown that the low-energy
effective theory including only the $d=5$ and $d=6$ operators contains an equal 
(a greater) number of real and imaginary parameters as the high-energy 
Seesaw model, when the number of right-handed neutrinos in the Seesaw theory 
is equal to (less than) the number of 
generations of Standard Model fermions.  Thus,  the determination of all  
$d=5$ and $d=6$ operator coefficients above would suffice {\it a priori} to determine all of 
the parameters of the high-energy Seesaw theory. In consequence, for instance, the leptogenesis rate can be written exclusively in terms of both operator coefficients~\cite{BGJ}. 
Other $d=6$ operators will be also present in the low-energy 
Lagrangian, since they are generated by radiative mixing of the above $d=6$ operator
in the renormalization
group running between the high-energy and  
low-energy scales.  The effects of these other $d=6$ operators
 are in consequence subdominant \cite{BGJ} and will not be further considered. The same statement will hold for all Seesaw theories considered in this work.
 
 \subsection{Scalar triplets: Type II Seesaw}
Assume now that the minimal SM matter content is enlarged only by the addition of a $SU(2)$ triplet of scalar fields $\overrightarrow \Delta$ with hypercharge $2$,
\begin{equation}
\overrightarrow \Delta=(\Delta_1, \Delta_2, \Delta_3)\,,
\end{equation}
whose relation to the physical charge eigenstates,  
\begin{equation}
(\Delta^{++}\,,\,\Delta^{+}\,,\,
\Delta^{0}\,)\,,
\end{equation}
is given by
\begin{equation}
\Delta^{++}\equiv\frac{1}{\sqrt{2}}(\Delta^1-i\Delta^2)\,,\quad\,\Delta^{+}\equiv\Delta^{3}\,,\quad\,\Delta^{0}\equiv\frac{1}{\sqrt{2}}(\Delta^1+i\Delta^2)\,.
\end{equation}
In the minimal Lagrangian,  gauge invariance allows a Yukawa coupling of the scalar triplet to two lepton doublets,
\begin{eqnarray}
\mathcal{L}^{\Delta\, ,\nu}_{{Y}} &=& 
\overline{\widetilde{\ell_\mathrm{L}}}\,Y_\Delta(\overrightarrow{\tau} \cdot \overrightarrow{\Delta})\,\ell_\mathrm{L} + {\rm h.c}.\,,
\end{eqnarray}
as well as 
a coupling of the scalar triplet to the Higgs doublet, 
\begin{eqnarray}
{\mu_\Delta}{\widetilde{\phi}}^{\dagger}(\overrightarrow{\tau}\cdot{\overrightarrow\Delta})^\dagger\phi
 + \text{h.c.}\; .
\end{eqnarray}
In these equations $\tau_i$ are the Pauli matrices, $Y_\Delta$ is a symmetric matrix in generation space and $\widetilde{\ell_\mathrm{L}}= i \tau_2(\ell_\mathrm{L})^c$ (i.e. $\overline{\widetilde{\ell_\mathrm{L}}}=-l^T_L C i \tau_2$).
  The minimal Lagrangian then writes: 
 \begin{eqnarray}
\!\!\mathcal{L}_\Delta\! \!\!& = &\!\! \!\!\left(D_{\mu}{\overrightarrow\Delta}\right)^{\dagger}\left(D^{\mu} {\overrightarrow\Delta}\right)+\left(\overline{\widetilde{\ell_\mathrm{L}}}Y_\Delta(\overrightarrow{\tau}\cdot{\overrightarrow\Delta})\ell_\mathrm{L}
+{\mu_\Delta}{\widetilde{\phi}}^{\dagger}(\overrightarrow{\tau}\cdot{\overrightarrow\Delta})^\dagger\phi
+ \text{h.c.}\right) \label{ScalarL}\\
\! \!\!&-& \!\!\!\!\!\left\{ {\overrightarrow\Delta}^{\dagger}{M_\Delta}^{2}{\overrightarrow\Delta}+\frac{1}{2}\lambda_{2}\left(\overrightarrow\Delta^{\dagger}\overrightarrow\Delta\right)^{2}+\lambda_{3}\left(\phi^{\dagger}\phi\right)\left(\overrightarrow\Delta^{\dagger}\overrightarrow\Delta\right)+\frac{\lambda_{4}}{2}\left(\overrightarrow\Delta^{\dagger}T^i\overrightarrow\Delta\right)^{2}+\lambda_{5}\left(\overrightarrow\Delta^{\dagger}T^i\overrightarrow\Delta\right)\phi^{\dagger}\tau^i\phi\right\} \nonumber\,,
\end{eqnarray}
where summation over the $SU(2)$ indices $i$ is assumed.
We choose to work in a basis in which $M_\Delta$ is real and diagonal and the covariant derivative  $D_{\mu}$ in Eq.~(\ref{ScalarL}) is given by
\begin{eqnarray}
\label{covder}
D_{\mu}\equiv\partial_{\mu}-ig\overrightarrow{T} \overrightarrow{W_{\mu}}-ig'B_{\mu}\frac{Y}{2}\, ,
\end{eqnarray}
 with $\overrightarrow{T}$ denoting the dimension-three representations of the $SU(2)$ generators,
\begin{eqnarray}
\label{generd3}
{T}_{1}=\left(\begin{array}{ccc}
0 & 0 & 0\\
0 & 0 & -i\\
0 & i & 0\end{array}\right)\:\:,\:\:{T}_{2}=\left(\begin{array}{ccc}
0 & 0 & i\\
0 & 0 & 0\\
-i & 0 & 0\end{array}\right)\:\:,\:\:{T}_{3}=\left(\begin{array}{ccc}
0 & -i & 0\\
i & 0 & 0\\
0 & 0 & 0\end{array}\right)\,.
\end{eqnarray}
The Lagrangian expressed  in terms of the charge components of the $\overrightarrow \Delta$ field can be found below, in Eq.~(\ref{ScalarLcomponents}).
Consider the limit  in which the triplets are heavy, ${M_\Delta}\gg v$. 
To solve the equation of motion for  $\Delta^{\alpha}$  in Eq.~(\ref{ScalarL}) and find the dominant terms of the effective low-energy Lagrangian up to $d=6$ operators,  it suffices to solve the problem perturbatively in the 
 quartic couplings of $\overrightarrow\Delta$, $\lambda_2$ and $\lambda_4$. 
At  zero order, it results:
\begin{eqnarray}
       \Delta^{\alpha}=\left[\left(D_{\mu}\right)^{2}+\lambda_{5}\overrightarrow{T}\phi^{\dagger}\overrightarrow{\tau}\phi+\left({M_\Delta}^{2}+\lambda_{3}\left(\phi^{\dagger}\phi\right)\right).{\bf1}_{\rm{isospin}}\right]_{\alpha\beta}^{-1}\left[{\mu^*_\Delta}\widetilde{\phi}^{\dagger}\tau^{\beta}\phi+\overline{\ell_\mathrm{L}}Y^{\dagger}_\Delta\tau^{\beta}\widetilde{\ell_\mathrm{L}}\right]\,.
\label{eqofmotion}
\end{eqnarray}
\subsubsection{Dimension 4 and 5 operators}
Expanding now the effective Lagrangian - using Eq.~(\ref{eqofmotion}) - in inverse powers of ${M_\Delta}$, one dimension four operator emerges:
\begin{eqnarray}
\label{d4}
\delta\mathcal{L}^{d=4}=\frac{|\mu_\Delta|^{2}}{M_\Delta^{2}}\left(\widetilde{\phi}^{\dagger}\overrightarrow{\tau}\phi\right)\left(\phi^{\dagger}\overrightarrow{\tau}\widetilde{\phi}\right)=2\frac{|\mu_\Delta|^2}{M_\Delta^{2}}\left(\phi^{\dagger}\phi\right)^{2}\,.
\end{eqnarray}
 We also obtain $\delta\mathcal{L}^{d=5 }$ as given in Eq.~(\ref{d=5}), with operator coefficients
given by the matrix 
\be\label{cd5scalar}
c^{d=5} = 4Y_\Delta\, \frac{ \mu_\Delta} {  M_\Delta^2} \,,
\ee  
which  at low energies leads to a light neutrino Majorana mass matrix of the form  
\be
m_\nu=  -2Y_\Delta v^2\,{\mu_\Delta \over { M_\Delta^2}} \,.
\label{mnutscal}
\ee
Notice that  neutrino masses turn out to be proportional to both $Y_\Delta$ and ${\mu_\Delta}$,  see Fig.~1. This is  as expected from the Lagrangian, Eq.~(\ref{ScalarL}), where the breaking of lepton number  symmetry L  results   precisely from the simultaneous presence of the Yukawa  and  $\mu_\Delta$ couplings\footnote{In the language of the full theory, this mass results when the neutral component of 
$\Delta$ acquires a vev 
$ <\Delta^0 >\equiv u/\sqrt{2}= \mu_\Delta v^2/ (\sqrt{2} M_\Delta^2)$, leading to a  Majorana mass matrix for the SM neutrinos, $m_\nu = -2 Y_{\Delta}\, u$. }  .
  It is important that, unlike for the fermionic Seesaw theories, the light neutrino mass matrix in Eq.~(\ref{mnutscal}) is only {\it linearly} dependent on the Yukawa coupling $Y_\Delta$. This means that the putative determination of the $d=5$ operator coefficients gives a direct access to the fundamental parameters $Y_\Delta$ of the high-energy theory, up to an overall scale $\mu_\Delta/M_\Delta^2$. We will analyze in Section \ref{pheno} the experimental access to  $\mu_\Delta/M_\Delta^2$ and to the elements of $Y_\Delta$.
 
\subsubsection{Dimension 6 operators}
 From Eq.~(\ref{eqofmotion}), the $d=6$ effective Lagrangian can also be obtained,  
\be\label{leff6scalar}
\delta {{\cal L}_\Delta}^{d=6} =  \delta {{\cal L}_{4F}} \,+\, \delta {{\cal L}_{\phi D}} \,+\, \delta {{\cal L}_{6\phi}} \,,
\ee
where 
\begin{eqnarray}
\left\{\begin{array}{l}
      \delta {{\cal L}_{4F}}=\frac{1}{{M_\Delta}^{2}}\left(\overline{\widetilde{\ell_\mathrm{L}}}\,  Y_\Delta\,\overrightarrow{\tau}\ell_\mathrm{L}\right)\left(\overline{\ell_\mathrm{L}}\overrightarrow{\tau}\,Y_\Delta^\dagger\,  \,\widetilde{\ell_\mathrm{L}}\right)\label{L4F} \\
       \delta {{\cal L}_{6\phi}} =-2\left(\lambda_{3}+\lambda_{5}\right)\frac{|\mu_\Delta|^{2}}{M_\Delta^{4}}\left(\phi^{\dagger}\phi\right)^{3}\label{L6fi}\\
      \delta {{\cal L}_{\phi D}}=\frac{|\mu_\Delta|^{2}}{M_\Delta^{4}}\left(\phi^{\dagger}\overrightarrow{\tau}\widetilde{\phi}\right)\left(\overleftarrow{D_\mu}\overrightarrow{D^\mu}\right)\left(\widetilde{\phi}^{\dagger}\overrightarrow{\tau}\phi\right)\label{LfiD}
   \end{array}\right.\,,
\end{eqnarray}
with the covariant derivative expressed in terms of  $(3\times3)$ $SU(2)$ generators, as in Eq.~(\ref{covder})\footnote{The first of these operators has already been derived in~\cite{Chao:2006ye}.}. Two of these operators can be rewritten in a more familiar form. After Fierz transformation, $\delta {{\cal L}_{4F}}$  can be expressed as
\begin{equation}
\label{L4Ffamiliar}
\delta {{\cal L}_{4F}}= 
-\frac{1}{M_\Delta^2}{Y_\Delta}_{ij}{Y_\Delta}^{\dagger}_{\alpha \beta}\left({\overline{\ell_L} _{\beta}}\gamma_\mu {\ell_L}_i \right)\left({\overline{\ell_L} _{\alpha}}\gamma_\mu {\ell_L}_{j }\right)\,,
\end{equation}
while the last operator in Eq.~(\ref{LfiD}) can  be recast as a combination of other operators which have been extensively studied in the literature (e.g. \cite{Buchmuller:1985jz}),
\begin{eqnarray}
\delta {{\cal L}_{\phi D}} 
&=&4\frac{|\mu_\Delta|^{2}}{M_\Delta^{4}}\left(\phi^{\dagger}\phi\right)\left[\left(D_{{\mu}}\phi\right)^{\dagger}\left(D_{{\mu}}\phi\right)\right]+4\frac{|\mu_\Delta|^{2}}{M_\Delta^{4}}\left[\phi^{\dagger}D_{{\mu}}\phi\right]^{\dagger}\left[\phi^{\dagger}D_{{\mu}}\phi\right]\,,\label{LfiD1}
 \end{eqnarray}
where the covariant derivative  is meant to be expressed in terms of Pauli matrices, 
\begin{eqnarray}
D_{\mu}=\partial_{\mu}-ig\frac{\tau_{a}}{2}W_{a\mu}-ig'B_{\mu}\frac{Y}{2}\,.
\end{eqnarray}
\subsubsection{Renormalization scheme}
 Four parameters of the SM are relevant to our discussions (in addition to fermion masses): $g,g^\prime, v$ and $\lambda$, the latter denoting the quartic self-coupling of the Higgs field,
\begin{eqnarray}
V=-\mu_\phi^{2}\left|\phi\right|^{2}+\lambda\left|\phi\right|^{4}\,.
\end{eqnarray}
To constrain the first three parameters,  we will work in the Z-scheme~\cite{DGHM}, that is, we will use as input  parameters the very-well determined experimental values of the fine structure constant $\alpha$ - as determined from Thompson scattering\footnote{An even more precise determination is now available from $g-2$ of the electron~\cite{PDG}.} -, the Fermi constant $G_F$ - as extracted from the muon decay rate by the removal of SM process-dependent radiative corrections -,  and the very precise measurement  of $M_Z$~\cite{PDG}.
The value of $\alpha$ is not affected by the presence of a scalar triplet, unlike the other parameters. 
$M_Z$ gets a correction from
$\delta {{\cal L}_{\phi D}}$ in Eq.~(\ref{LfiD})
\begin{eqnarray}
\frac{\delta M_{Z}^{2}}{M_{Z}^{2}}=2v^{2}\frac{|\mu_\Delta|^{2}}{M_\Delta^{4}}\,.
\end{eqnarray}
Similarly, the 4-fermion operator $\delta {{\cal L}_{4F}}$ affects the extraction of the value of the Fermi constant from muon decay. Defining, as it is customary, this constant as the coefficient  in 
\begin{equation}
-\frac{4 G_F}{\sqrt{2}}\left({\overline{\ell_L} _{\nu_\beta}}\gamma_\mu {\ell_L}_\mu\right)\left({\overline{\ell_L} _{e}}\gamma_\mu {\ell_L}_{\nu_\alpha}\right)\,,
\end{equation}
it is easily seen that $\delta {{\cal L}_{4F}}$ in Eq.~(\ref{L4Ffamiliar})
    induces in turn a  shift with respect to the ``Standard Model definition" $G_F^{SM}=1/(\sqrt{2}v^2)$\footnote{Note that with a scalar triplet $\frac{1}{2v^2}\neq \frac{g^2}{8 M_W^2}$ due to the scalar triplet induced $M_W$ shift, see below.},  which affects the value extracted from muon decay,
\begin{eqnarray}
\label{G_F}
\delta G_{F}=
 \frac{1}{ \sqrt{2} M_\Delta^2}
  |Y_{\Delta_{e \mu}}|^2 
\end{eqnarray}
\begin{eqnarray}
G_F= G_F^{SM}\,+\,\delta G_F\,.
\label{GFMscalar}
\end{eqnarray}
 The quartic self-coupling of the Higgs field is also renormalized 
by the dimension four operator obtained in the effective theory, Eq.~(\ref{d4}),  
\begin{eqnarray}
\delta \lambda= -2\frac{|\mu_{\Delta}|^2}{M_{\Delta}^{2}}\,, 
\end{eqnarray}
influencing the location of the minimum of the Higgs potential. 
Another  $d=6$ operators, $\delta {{\cal L}_{6\phi}}$ in Eq.~(\ref{L6fi}), also modifies the Higgs potential, which all in all becomes
\begin{eqnarray}
V=-\mu_\phi^{2}\left|\phi\right|^{2}+(\lambda+\delta\lambda)\left|\phi\right|^{4}+2\left(\lambda_{3}+\lambda_{5}\right)\frac{|\mu_\Delta|^{2}}{M_\Delta^{4}}\left|\phi\right|^{6}\,,
\end{eqnarray}
 inducing a shift in the vacuum expectation value of the Higgs field, 
\begin{eqnarray}
\frac{\delta v^2}{v^2}=-{3} v^2 \frac{|\mu_\Delta|^{2}}{M_\Delta^{4}}\frac{(\lambda_{3}+\lambda_{5})}{\lambda + \delta \lambda}\,.
\end{eqnarray}
 Using all these renormalized parameters, in Sect. \ref{typeIIpheno} we will consider the deviations - with respect to the SM predictions - induced by the new physics on the values taken by a variety of physical observables. 
 
 Finally, as regards the relative number of parameters in the high and low energy theories, 
 the inclusion in the latter of only the $d=5$ and $d=6$ operators above does not 
 suffice to match the number of free parameters of the full scalar-triplet Seesaw theory, as can be easily 
 deduced from the comparison of Eq.~(\ref{ScalarL}) with Eqs. (\ref{leff6scalar}) and (\ref{LfiD}). Up to $d=8$ operators would have to be considered for this purpose, which is beyond the scope of the present work.
\subsection{Fermionic triplets: Type III Seesaw}
\label{fermtriptheory}
Consider now the SM field content extended by the only addition of fermions which are triplets of $SU(2)$ with zero hypercharge, hereafter denoted by $\vec \Sigma$, where the vectorial character refers to its three $SU (2)$-components, $\vec \Sigma=(\Sigma^1,\Sigma^2,\Sigma^3)$.  Being $\vec \Sigma$ in the adjoint representation of the gauge group, its Majorana mass term is gauge invariant and the interactions are described by the Lagrangian
\begin{eqnarray}
\label{Lfermtrip}
{\mathcal L}_{\Sigma} &=& 
i\,\overline{\vec{\Sigma}_{R}} 
\D\,
\vec{\Sigma}_R
-\large[\,\frac{1}{2} \overline{\vec \Sigma_{R}} M_\Sigma \vec \Sigma_{R}^c
+\overline{\vec \Sigma_{R}} Y_{\Sigma}(\widetilde \phi^\dagger \vec \tau \ell_{L}) 
+\text{h.c.}\,\large]\; .
\end{eqnarray} 
In this equation, the covariant derivative is given by Eqs.~(\ref{covder})  and (\ref{generd3}) and the three $SU (2)$-components of the field $\vec \Sigma$
 have (identical) Majorana mass terms.  They are not eigenstates of the electric charge, which would be given instead by the combinations
 \begin{eqnarray}
 \Sigma^{\pm}\equiv\frac{\Sigma^{1}\mp i\Sigma^{2}}{\sqrt{2}}\,\,\,,\,\,\,\,\Sigma^{0}\equiv\Sigma^{3}\,.
 \end{eqnarray}
We will work throughout in a basis in which $M_\Sigma$ is  a diagonal 
matrix in generation space. 
 The Yukawa coupling $Y_\Sigma$ in Eq.~(\ref{Lfermtrip}) is then a general matrix in generation space. After electroweak symmetry breaking, this term induces  Majorana neutrino masses for the left-handed neutrino fields of the SM through the exchange of $\vec \Sigma$ particles, see Fig.~\ref{trois-types}.
 
\subsubsection{Dimension 5 operator}
 Solving the equations of motion, it results that 
 \begin{eqnarray}
{\vec{\Sigma}_{R}} &=&P_{R}\left[iD\!\!\!\!/-M_{\Sigma}\right]^{-1} \left[{Y_{\Sigma}^{*}}\phi^{\dagger}\vec{\tau} \widetilde{\ell_{L}}+{Y_{\Sigma}} \widetilde{\phi}^{\dagger}\vec{\tau} \ell_{L}\right]\nonumber\\
&=&-\frac{1}{M_{\Sigma}}{Y_{\Sigma}^{*}}\phi^{\dagger}\vec{\tau} 
\widetilde{\ell_{L}}-\frac{1}{M_\Sigma^\dagger}{iD\!\!\!\!/}\frac{1}{M_\Sigma}{Y_{\Sigma}}\widetilde{\phi}^{\dagger}\vec{\tau} \ell_{L}+\mathcal{O}\left(\frac{1}{M_{\Sigma}^{3}}\right)\,,
\end{eqnarray} 
where $i,j$ are $SU(2)$ indices, $i,j=1,2,3$.
 This allows to obtain the $d=5$ operator in Eq.~(\ref{d=5}), with coefficient matrix given in this case by 
 \be\label{cd5ft}
c^{d=5} =   Y_\Sigma^T \,\frac{1}{ M_\Sigma} \,Y_\Sigma\;,
\ee  
which leads at low energies to a light neutrino Majorana mass matrix of the form  
\be
m_\nu= -\frac{v^2}{2} \, Y_\Sigma^T \,\frac{1}{ M_\Sigma} \,Y_\Sigma\;.\label{mnuft}
\ee
\subsubsection{Dimension 6 operator}
At the next order in the effective Lagrangian, we obtain a unique operator\footnote{ We thank S. Antusch for helping to clarify the derivation of this operator in an early stage.} :
\be\label{cd6fta}
\delta{\cal L}^{d=6} = c^{d=6}_{\alpha \beta} \, \left( \overline{\ell_{L\alpha}} \vec\tau \tilde \phi
\right) i \D \left( \tilde \phi^\dagger \vec\tau \ell_{L \beta} \right),
\ee 
where the $d=6$ operator coefficients are given in terms of the parameters
of the high-energy Seesaw theory by
\be\label{cd6ft}
c^{d=6} = Y_\Sigma^\dagger \, \frac{1}{M_\Sigma^\dagger}\frac{1}{ M_\Sigma} \, Y_\Sigma \, .
\ee  
Notice the large parallelism between the results for this Seesaw scenario mediated by fermionic triplets and those for the minimal Seesaw based on the exchange of fermionic singlets, Eqs.~(\ref{cd5}) and (\ref{d6})-(\ref{cd6}). The main difference is that, now,  in the $d=6$ operator in Eq.~(\ref{cd6fta}) the  interaction terms in the covariant derivative are active,
 as the quantities in  brackets are $SU(2)$ triplets, which amounts to a richer interaction pattern.  
 
  A first consequence is that, when the Higgs doublet acquires a vacuum
expectation value, the $d=6$ operator  corrects both the $d=4$ kinetic energy terms of light leptons {\it and} their couplings to W bosons, while no corrections to the hypercharge boson $B_\mu$ appeared, because the combinations in brackets in Eq.~(\ref{cd6fta}) have zero hypercharge.
  After electroweak symmetry breaking, the part of the effective Lagrangian concerning leptons is, in the flavour basis, 
\begin{eqnarray}
&\,&\mathcal{L}^{d\leq6}_{\textrm{leptons}}=i\overline{\nu_{L}}_{\alpha}\partial\!\!\!/ \left(\delta_{\alpha\beta}+\epsilon^{\Sigma}_{\alpha\beta}\right){\nu_{L}}_{\beta}+i\overline{l_{L}}_{\alpha}\partial\!\!\!/ \left(\delta_{\alpha\beta}+2\epsilon^{\Sigma}_{\alpha\beta}\right){l_{L}}_{\beta} +i\overline{l_{R}}_{\alpha}\partial\!\!\!/ {l_{R}}_{\alpha}\nonumber\\
& &
-\frac{1}{2}\left[
\overline{{\nu_L}^c}_\alpha\,m_{\nu\,\alpha\beta}\,{\nu_L}_\beta + \text{h.c.}\right]
-\left[\overline{l_R}_\alpha\,m_{l\,\alpha\beta}\,{l_L}_\beta + \text{h.c.}\right] 
+\frac{1}{\sqrt{2}}g\left[\overline{l_{L}}_{\alpha}{W\!\!\!\!\!/}\;^{-}\left(\delta_{\alpha\beta}+2\epsilon^{\Sigma}_{\alpha\beta}\right){\nu_{L}}_{\beta}+\text{h.c.}\right]\nonumber
\\
& &-\frac{g}{2}\overline{l_{L}}_{\alpha}{W\!\!\!\!\!/}\;^{3}\left(\delta_{\alpha\beta}+4\epsilon^{\Sigma}_{\alpha\beta}\right){l_{L}}_{\beta}+\frac{g}{2}\overline{\nu_{L}}_{\alpha}{W\!\!\!\!\!/}\;^{3}{\nu_{L}}_{\alpha}-\frac{g'}{2}\overline{l_{L}}_{\alpha}{B\!\!\!\!/}\;{l_{L}}_{\beta}-\frac{g'}{2}\overline{\nu_{L}}_{\alpha}{B\!\!\!\!/}\;{\nu_{L}}_{\alpha}\,,
\end{eqnarray}
where 
\begin{eqnarray}
\label{epsilonfermtrip}
\epsilon^{\Sigma}\equiv\frac{v^{2}}{2}c^{d=6}\,,
\end{eqnarray}
 with $c^{d=6}$ as defined in Eq.~(\ref{cd6ft}) and $m_l$ denoting the charged
   lepton mass matrix. We assume hereafter a choice of basis in which both $m_l$ and $M_\Sigma$ are diagonal.
The neutrino {\it and} charged lepton fields need now to be normalized in order to acquire canonically normalized kinetic terms. At order $1/M^2$, i.e. linear in 
the parameters $\epsilon^\Sigma_{\alpha \beta}$, the redefinitions
\begin{eqnarray}
{\nu_{L}}_{\alpha}&\rightarrow&{\nu'_{L}}_{\alpha}\equiv \left(\delta_{\alpha\beta}+\frac{1}{2}\epsilon^{\Sigma}_{\alpha\beta}\right){\nu_{L}}_{\beta}\,,\nonumber\\
{l_{L}}_{\alpha}&\rightarrow&{l'_{L}}_{\alpha}\equiv \left(\delta_{\alpha\beta}+\epsilon^{\Sigma}_{\alpha\beta}\right){l_{L}}_{\beta}\,,
\end{eqnarray}
results in a Lagrangian in the flavour basis which, at order ${\cal O}(1/M^2)$, takes the form (primes on the fields will be disregarded),
\begin{eqnarray}
\label{LagShift2}
\mathcal{L}^{d\leq6}_{\textrm{leptons}} &=& i\overline{\nu_{L}}_{\alpha}\partial\!\!\!/ {\nu_{L}}_{\alpha}+i\overline{l_{L}}_{\alpha}\partial\!\!\!/{l_{L}}_{\alpha} +i\overline{l_{R}}_{\alpha}\partial\!\!\!/ {l_{R}}_{\alpha}
-\frac{1}{2}
\left[\overline{{\nu_L}^c}_\alpha\,{m^\prime_\nu}_{ \alpha\beta}\,{\nu_L}_\beta
+ \text{h.c.}\right] \nonumber\\
&-&\left[\overline{l_R}_\alpha\,m^\prime_{l\,\alpha\beta}\,{l_L}_\beta + \text{h.c.}\right] 
+ \mathcal{L}_{CC} + \mathcal{L}_{NC}+\mathcal{L}_{em}\,,
\end{eqnarray}
where $m^\prime_\nu\equiv (1-\epsilon^*/2)m_\nu (1-\epsilon/2)$, $m^\prime_l \equiv m_l (1-\epsilon)$ and 
\begin{eqnarray}
\mathcal{L}_{CC} &=&\frac{g}{\sqrt{2}}\overline{l_{L}}_{\alpha}{W\!\!\!\!\!/}\;^{-}\left(\delta_{\alpha\beta}+\frac{1}{2}\epsilon^{\Sigma}_{\alpha\beta}\right){\nu_{L}}_{\beta}+ \text{h.c.}\,,\label{LagCCNCfermtrip}
\\
\mathcal{L}_{NC}&=&\frac{g}{cos\theta_{W}}\left\{\frac{1}{2}\left[\overline{\nu_{L}}_{\alpha}\gamma_{\mu}\left(\delta_{\alpha\beta}-\epsilon^{\Sigma}_{\alpha\beta}\right){\nu_{L}}_{\beta}-\overline{l_{L}}_{\alpha}\gamma_{\mu}\left(\delta_{\alpha\beta}+2\epsilon^{\Sigma}_{\alpha\beta}\right){l_{L}}_{\beta}\right]-sin^{2}\theta_{W}J^{em}_{\mu}\right\}Z^{\mu}\nonumber \,,\\
\mathcal{L}_{em}&=&eJ^{em}_{\mu}A^{\mu}\nonumber\,,
\end{eqnarray}
where $J^{em}_{\mu}=-\overline{l}\gamma_\mu l$ is the electromagnetic current. 
We can finally rotate to the basis in which both the lepton kinetic energies and 
their mass matrices are diagonalized (for details see Appendix A), 
\bea
\label{fermtripLm}
{\cal L}_{\rm leptons}^{d\le 6}=
\frac{1}{2}\overline {\nu_i} \left( i\dv- m^{diag}_{\nu\,i} \right) {\nu_i}+\frac{1}{2}\overline {l_i} \left( i\dv- m^{diag}_{l\, i} \right) {l_i}+ \mathcal{L}_{CC} + \mathcal{L}_{NC}+\mathcal{L}_{em}\,.
\eea
A non-unitary mixing matrix $N$ replaces now the usual unitary $U_{PMNS}$ matrix in the charged current couplings contained in Eq.~(\ref{fermtripLm}),
because of the flavour-dependent field rescaling involved,  while the couplings to the $Z$ boson acquire also a non-unitary mixing pattern,
\bea
\label{JCC_fermtrip}
J_\mu^{-\,CC}  
&\equiv& \overline {l_L} \, \gamma_\mu \,
N \, \nu,\\
\label{JNCnu_fermtrip}
J_\mu^{3} (\text{neutrinos})
&\equiv& {1 \over 2} \overline \nu \,\gamma_\mu
(N^\dagger \, N)^{-1}
\,\nu\,,\\
\label{JNCe_fermtrip}
J_\mu^{3} (\text{leptons})
&\equiv& {1 \over 2} \overline l \,\gamma_\mu
(NN^\dagger)^2
\,l .
\eea
The non-unitary mixing matrix $N$ is a function of the $d=6$ coefficient matrix 
\bea
\label{NSigma}
N\equiv \Omega\,{U^l_L}^\dagger \left(1+\frac{1}{2}\epsilon^{\Sigma}\right)\,U^\nu\,,
\eea
where, once again, $\Omega\equiv{\rm diag}(e^{i\omega_1},e^{i\omega_2},e^{i\omega_3})$ reabsorbs three unphysical phases in the definition of the charged lepton fields, and the matrices $U^\nu$ and $U^l_L$ diagonalize the effective leptonic mass matrices\footnote{ Within the order $\epsilon^\Sigma$ used throughout, the mass eigenvalues are defined at first order in it and thus the eigenvectors should be consistently defined at order zero in that expansion. As a consequence, any representation of the leptonic matrices $U^l_{L, R}$ which diagonalizes the mass matrix has to be physically equivalent to the identity.}, $m_\nu^\text{diag}\equiv {U^\nu}^T \, m_\nu\,U^\nu\,,
  m_l^\text{diag}\equiv {U^l_R}^\dagger \, m_l\, (1 - \epsilon)\,U^l_L\,$ (see Appendix A).   When the flavour basis chosen is such that  both $U^l_L$ and $\Omega$ are equal to the identity matrix, and taking into account that $U^\nu$ does not receive corrections from $c^{d=6}$ at  ${\cal O}(1/M_\Sigma^2)$ , $N$ simplifies to
   \bea
\label{NSigmabasis}
N\equiv \left(1+\frac{1}{2}\epsilon^{\Sigma}\right)\,U_{PMNS}\,
\eea
and, consequently, $NN^\dagger=1+\epsilon^\Sigma$, $N^\dagger N= U_{PMNS}^\dagger(1+\epsilon^\Sigma) U_{PMNS}$.
These expressions can be compared with the equivalent ones for the singlet-fermion Seesaw theory, Eq.~(\ref{NNbasis}) and below it.
Whatever the basis, the currents in Eqs.~(\ref{JCC_fermtrip})-(\ref{JNCe_fermtrip})  can also be compared with the corresponding ones for 
 the singlet-fermion Seesaw theory,  Eq.~(\ref{JCC_d6}) and (\ref{JNC_d6}). {\it  A non-unitary mixing pattern has appeared in both cases, although the modified  $Z$-neutrino couplings differ and non-unitary flavour mixing is now also present in the Z-charged lepton couplings}.

  An important consequence of the flavour-changing W- and Z-lepton couplings is their  contribution to muon decay into electron plus missing energy, which modifies the definition of $G_F$ as extracted from muon decay, as follows:
\be
\label{G_F_fermtrip}
G_F =
G_F^{SM}
\sqrt{(NN^\dagger)_{ee} (NN^\dagger)_{\mu\mu} + {3\over 4} [(NN^\dagger)^2_{e \mu }]^2 
}\sim G_F^{SM}
\sqrt{(NN^\dagger)_{ee} (NN^\dagger)_{\mu\mu} \,,
}
\ee
where higher order  correction, ${\cal O}((\epsilon^\Sigma)^2)$,  have been neglected in the last step.
Its phenomenological consequences will be explored in Sect.~(\ref{sectfermtrip}).
 
  Finally, in analogy with the case of the fermionic singlet Seesaw theory, it is remarkable that departures from unitarity of the leptonic mixing matrix can be now directly related to the $d=6$ operator coefficients and thus to combinations of the high-energy parameters,
\begin{equation}
\label{Nc6-ft}
|NN^\dagger -1|_{\alpha\beta} \,=\,|\epsilon^\Sigma|\,=\,\frac{v^2}{2}\,|c^{d=6}|_{\alpha\beta}\,=\,\frac{v^2}{2}\,|Y_\Sigma^\dagger\frac{1}{M_\Sigma^\dagger}\frac{1}{ M_\Sigma}Y_\Sigma|_{\alpha\beta}\,.
\end{equation}
Once again, in the flavour basis in which $\Omega$ and $U^l_L$ equal the indentity matrix, the absolute-value bars in this equation can be dropped. 
In Sect.~\ref{sectfermtrip} the present numerical constraints on $|c^{d=6}|$ will be explored.
 
\subsubsection{Parameter counting}
 Finally, it can be shown that the low-energy
effective theory, including only the $d=5$ and $d=6$ operators,  contains in this case an equal 
(greater) number of real and imaginary parameters as the high-energy 
Seesaw model,  when the number of right-handed fermionic triplet generations in the Seesaw theory 
is equal to (less than) the number of 
generations of Standard Model fermions.  The demonstration is equivalent to that in Ref.~\cite{BGJ} for the case of singlet-fermion Seesaw theory. The kinetic energy terms in the Lagrangian, Eq.~(\ref{Lfermtrip}), 
 are invariant under the chiral transformations
\begin{eqnarray}
\ell_{L}&\rightarrow& V_{\ell}\ell_{L}\,,\nonumber\\
e_R &\rightarrow& V_{e}e_{R}\,,\\
\Sigma_{R} &\rightarrow & V_{\Sigma}\Sigma_{R}\nonumber\,,
\end{eqnarray}
where the $V$'s are unitary transformations. Consider first the complete theory with 
 $n$ lepton families and $n'$ right-handed fermionic triplets. The Yukawa terms and the Majorana mass term are not invariant under such chiral symmetry,  but invariance can be recovered if they are considered as spurion fields  transforming as
\begin{eqnarray}
Y_{e} &\rightarrow& Y'_{e} \equiv V_{\ell}Y_{e} V_{e}^{\dagger}\,,\nonumber\\
Y_{\Sigma} &\rightarrow& Y'_{\Sigma} \equiv V_{\Sigma}Y_{\Sigma}V_{\ell}^{\dagger}\,,\\
M_{\Sigma} &\rightarrow& M'_{\Sigma} \equiv V_{\Sigma}M_{\Sigma}V_{\Sigma}^{t}\,.\nonumber
\end{eqnarray}
Counting how many physical parameters $N_{phys}$ are needed to describe the Yukawa
and Majorana mass
terms in the Seesaw Lagrangian is tantamount to counting how many
equivalence classes there exist with respect to these transformations.  
The result is given by 
\be
\label{nphys}
N_{phys}= N_{\rm order}-(N_G-N_{H}),
\ee
where $N_{\rm order}$ is the total number of parameters contained in
the Yukawa and Majorana mass matrices, $N_{G}$ is the number of parameters contained in
the matrices of the 
chiral symmetry group $G=U(n)_\ell \times U(n)_e \times U(n')_N$.  $N_{H}$ is
the number of parameters contained in the matrices of the subgroup $H$ of 
the chiral symmetry group which remains unbroken by the Yukawa and Majorana mass matrices: in  the present model there is no unbroken subgroup $H$ because of lepton number violation. Table~\ref{parametersIIIfull} summarizes the result for the high-energy theory.\begin{table}[ht]
\begin{center} 
 Seesaw Model
\end{center}
\begin{center}
\begin{tabular}{|c|cc|}
\hline
Matrix &Moduli &Phases \\
\hline
$Y_e$ &$n\times n$ &$n\times n$\\
$Y_\Sigma$ &$n\times n'$ &$n\times n'$\\
$M_{\Sigma}$ &$\frac{n'(n'+1)}{2}$ &$\frac{n'(n'+1)}{2}$ \rule[-8pt]{0pt}{23pt}\\
\hline
$V_e$ &$\frac{n(n-1)}{2}$ &$\frac{n(n+1)}{2}$\rule[-8pt]{0pt}{23pt}\\
$V_\ell$&$\frac{n(n-1)}{2}$&$\frac{n(n+1)}{2}$\rule[-8pt]{0pt}{23pt}\\
$V_{\Sigma}$ &$\frac{n'(n'-1)}{2}$ &$\frac{n'(n'+1)}{2}$\rule[-8pt]{0pt}{23pt}\\
\hline
$N_{phys}$ &$n+n'+ n n'$ &$n(n'-1)$ \\
\hline
\end{tabular}
\end{center}
\caption{Number of physical parameters, for $n$ light and $n^\prime$ heavy neutrino generations.}
\label{parametersIIIfull}
\end{table}
This is to be compared with the effective low-energy Lagrangian including operators of $d\le6$. It is invariant under the chiral transformations only if
\begin{eqnarray}
c^{d=5}&\rightarrow&V_{\ell}^{*}c^{d=5}V_{\ell}^{\dagger}\,,\nonumber\\
c^{d=6}&\rightarrow&V_{\ell}c^{d=6}V_{\ell}^{\dagger}\,.
\end{eqnarray}
$c^{d=5}$ is a complex symmetric matrix and $c^{d=6}$ is a complex hermitian matrix and since the dimension 5 operators breaks lepton number, there is no unbroken subgroup that remains. The corresponding counting of parameters is shown in Table~\ref{parametersIIIeff}, to be compared with that in Table~\ref{parametersIIIfull} for the high-energy theory.
\begin{table}[ht]
\begin{center}
 Effective Theory ($d\le 6$)
\end{center}
\begin{center}
\begin{tabular}{|c|cc|}
\hline
Matrix &Moduli &Phases \\
\hline
$Y_e$ &$n\times n$ &$n\times n$ \\
$c^{d=5}$ &$\frac{n(n+1)}{2}$ &$\frac{n(n+1)}{2}$\rule[-8pt]{0pt}{23pt} \\
$ c^{d=6}$ &$\frac{n(n+1)}{2}$ &$\frac{n(n-1)}{2}$ \rule[-8pt]{0pt}{23pt}\\
\hline
$V_e$ &$\frac{n(n-1)}{2}$ &$\frac{n(n+1)}{2}$\rule[-8pt]{0pt}{23pt}\\
$V_\ell$ &$\frac{n(n-1)}{2}$ &$\frac{n(n+1)}{2}$ \rule[-8pt]{0pt}{23pt}\\\hline
$N_{phys}$ &$n(n+2)$ &$n(n-1)$ \\
\hline
\end{tabular}
\end{center}
\caption{Number of physical parameters, for $n$ light lepton generations.}
\label{parametersIIIeff}
\end{table}
Thus,  the determination of all  
$d=5$ and $d=6$ operator coefficients above would again suffice {\it a priori} to determine all of 
the parameters of the high-energy Seesaw theory with two or three heavy neutrino generations. In consequence, for instance, the leptogenesis rate could be written exclusively in terms of both operator coefficients~\cite{inpreparation}.
\vspace{2cm}
\subsection{Summary}
  To conclude this Section, we have gathered in Table~\ref{tableops}
  the $d=6$ operators obtained for the three basic Seesaw scenarios,
  together with the corresponding expressions for the elements of the
  $d=6$ coefficient matrices. The elements of the $d=5$ coefficient
  matrices are included as well.
  
  \vspace{2cm}
 \begin{table}[!h]
\begin{center}
\begin{tabular}{|c||c|c|c|}
\hline
& \multicolumn{3}{c|}{Effective Lagrangian $\mathcal{L}_{eff}=c_{i}\mathcal{O}_{i} $} \rule[-5 pt]{0pt}{18 pt}\\
\cline{2-4}
Model & $c^{d=5}$ & $c^{d=6}_{i}$ & $\mathcal{O}^{d=6}_{i}$  \rule[-5 pt]{0pt}{18 pt}\\
\hline
\hline
Fermionic Singlet & $Y_{N}^{T}\frac{1}{M_{N}}Y_{N}$ & $\left(Y_{N}^{\dagger}\frac{1}{M_{N}^\dagger}\frac{1}{ M_N}Y_{N}\right)_{\alpha\beta}$ & $\left(\overline{\ell_{L\alpha}}\widetilde{\phi}\right)i\partial\!\!\!/\left(\widetilde{\phi}^{\dagger}\ell_{L\beta}\right)$ \rule[-14 pt]{0pt}{34 pt}\\
\hline
 & & $\frac{1}{ M_\Delta^{2}}Y_{\Delta \alpha\beta}Y_{\Delta \gamma\delta}^{\dagger}$ &$ \left(\overline{\widetilde{\ell_{L\alpha}}}\overrightarrow{\tau}\ell_{L\beta}\right)\left(\overline{\ell_{L\gamma}}\overrightarrow{\tau}\widetilde{\ell_{L\delta}}\right)$  \rule[-14 pt]{0pt}{34 pt}\\
\cline{3-4}
Scalar Triplet & $4Y_{\Delta}\frac{\mu_{\Delta}}{M_{\Delta}^{2}}$ & $\frac{|\mu_{\Delta}|^{2}}{M_{\Delta}^{4}}$ & $\left(\phi^{\dagger}\overrightarrow{\tau}\widetilde{\phi}\right)\left(\overleftarrow{D_{\mu}}\overrightarrow{D^{\mu}}\right)\left(\widetilde{\phi}^{\dagger}\overrightarrow{\tau}\phi\right) $ \rule[-14 pt]{0pt}{34 pt}\\
\cline{3-4}
  & & $-2\left(\lambda_{3}+\lambda_{5}\right)\frac{|\mu_\Delta|^{2}}{M_\Delta^{4}}$ & $\left(\phi^{\dagger}\phi\right)^{3}$   \rule[-10 pt]{0pt}{25 pt}\\
\hline
Fermionic Triplet & $Y_{\Sigma}^{T}\frac{1}{M_{\Sigma}}Y_{\Sigma}$ & $\left(Y_{\Sigma}^{\dagger}\frac{1}{M_{\Sigma}^\dagger}\frac{1}{ M_\Sigma}Y_{\Sigma}\right)_{\alpha\beta}$ & $\left(\overline{\ell_{L\alpha}}\overrightarrow{\tau}\widetilde{\phi}\right)iD\!\!\!\!/\left(\widetilde{\phi}^{\dagger}\overrightarrow{\tau}\ell_{L\beta}\right)$  \rule[-14 pt]{0pt}{34 pt}\\
\hline
\end{tabular}
\end{center}
\caption{ Coefficients of the $d=5$ operator, $c^{d=5}$, and  $d=6$ operators and their coefficients, $c^{d=6}$, in the three basic Seesaw theories.}
\label{tableops}
\end{table}
\newpage
\section{Low scale Seesaw $M\,\sim\,$$\mathcal{O}$$(TeV)$}
\label{lowscale}
\subsection{Electroweak Hierarchy problem}
If the Seesaw scale is far above the electroweak scale, the theory clashes  with the electroweak hierarchy problem, that is, the fact that data indicate a value for  
 the  Higgs mass of the order of the electroweak scale, $v\sim{\cal O} (100)$GeV. Such a mass is unnaturally light if there is new physics beyond the SM and at a higher scale, to which the Higgs boson is sensitive. The three minimal scenarios considered in the previous Section do face this  problem if the new scales are much larger than the electroweak scale $ v$.
 
  Indeed, for the Seesaw Type I, the one-loop contribution to the Higgs mass has been computed long ago~\cite{Casas:2004gh},
 \begin{eqnarray}
 \label{hierarchyI}
\delta m_H^2=-{Y_N^\dagger Y_N\over 16\pi^2}
\left[2\Lambda^2 + 
2M_N^2 \log{M_N^2\over \Lambda^2}\right]\,,
\end{eqnarray}
while in  the case of the scalar-triplet (type II), we find that the contribution is given by\footnote{No dependence on the quartic coupling $\lambda_5$ of the Lagrangian Eq.~(\ref{ScalarL}) appears, as the Higgs fields are combined in this term in a triplet of $SU(2)$, while 
the Higgs mass  is a singlet.}  
\begin{equation}
\label{hierarchyII}
\delta m^2_H=  \frac{1}{16 \pi^2} \left[
3 \lambda_3 (\Lambda^2-M^2_\Delta \log \frac{\Lambda^2}{M^2_ 
\Delta})
-12 |\mu_\Delta|^2 \log \frac{\Lambda^2}{M^2_\Delta}   \right]\,,
\end{equation}
and, finally, for the fermionic-triplet Seesaw (type III), we obtain 
\bea
\label{hierarchyIII}
\delta {m_H}^2= -3\,{Y_\Sigma^\dagger Y_\Sigma\over 16\pi^2}\,
\left[2\Lambda^2 + 
2M_\Sigma^2 \log{M_\Sigma^2\over \Lambda^2}\right]
\,,
\eea
where $\Lambda$ is the regulator cutoff. In these equations, terms proportional to $v^2$ and $m_H^2$ have been neglected. 
   Eqs.~(\ref{hierarchyI})-(\ref{hierarchyIII})   all show a quadratic sensitivity to the new scales characteristic of Seesaw theories, 
  implying  that large fine-tunings would be necessary to accommodate the experimental data if {\it any}  of the new  scales introduced is much larger than $v$ (or the Yukawa couplings are not extremely fine-tuned in Type I and III Seesaw). 
  
  For instance, imposing that the one-loop correction is not larger than the Higgs mass itself, let's say $m_H=150$~GeV for definiteness, $M_N$ and $M_\Sigma$ should be below $\sim 10^7$ GeV for Yukawa couplings of order $m_\nu^{1/2} M_{N,\Sigma}^{1/2}/v$, while $M_\Delta$ should be below a scale which depends on $\lambda_3$ and $\mu_\Delta$.
 In any case, for scales not much larger than the electroweak one, the contribution of the Seesaw theory to the hierarchy problem would be obviously avoided. As a by-product, new exciting physics signals would then be expected at present and future experimental facilities.
 
 The question we wish to analyze now is whether it is indeed possible that nature has chosen the high energy scale $M$ of the Seesaw scenario  close to the electroweak scale, rather than to the Grand Unified scale, with ${\cal O}(1)$ Yukawa couplings, without  fine-tuning the parameters and in particular the Yukawa couplings. 
 
 \subsection{Direct Lepton Violation}
After all, the analysis of the previous Sections has shown that, while neutrino masses result from a lepton-number odd $d=5$ operator, other manifestations of the new physics behind are encoded in lepton-number conserving  $d=6$ operators (as well as in higher dimensional operators). As lepton number appears to be an approximate symmetry of nature, it is natural to assume that it may be broken through small parameters - such as those responsible for neutrino masses -, while other beyond the SM effects  of the high-energy theory, which are lepton-number preserving, need not  be strongly suppressed. The choice of such a L-odd small parameter may be thus a natural one, as it corresponds to the breaking of a symmetry and its value cannot be destabilized by other large scales of the theory through radiative corrections, because by nature it can only be multiplicatively renormalized.
 
  Assume thus $M$ ($M_N$, $M_\Delta$, $M_\Sigma$) to be higher but not far from the TeV scale. The issue is then whether it is possible to decouple and further suppress the coefficients of the $d=5$ operators from those of the fermionic $d=6$ operators, without appealing to  fine-tunings and cancellations in the Yukawa parameters or heavy mass matrices~\footnote{Although operators of dimension higher than six are increasingly relevant as the scale is lowered toward the electroweak scale, an  analysis restrained to the $d=5$ and $d=6$ operators should still convey the main physical aspects, as long as the scale keeps being larger than ${\cal O}(v)$. }.
  If this is possible, the tiny values of the neutrino masses could be accommodated, while the effects of the $d=6$ operators - suppressed only as $1/M^2$ - would be close to observability~\footnote{Note that Ref.~\cite{Cirigliano:2005ck} studied the effects of various higher-order operators in a completely different context: the dissociation of flavour violation scale and lepton number violation scale in extended theories, while we focus on the dissociation of d=5 and d=6 operators characteristic of the minimal seesaw models.}.
    
As a guideline to achieve such a scenario recall that, because Majorana neutrino masses are forbidden in the SM, light neutrinos inheritate their Majorana character from a Majorana source in the high-energy theory. This implies that light Majorana neutrino masses have to vanish either when the new Majorana scale goes to infinity and the new physics decouples, or proportionally to it. 
  A quick look at Table~\ref{tableops}, together with the pattern of operator coefficients 
  found for the case of scalar-triplet mediated Seesaw mechanisms, suggests the following {\it ansatz}:
  
  {\it When the breaking of L symmetry takes place in the full theory through a small mass scale $\mu$, distinct from the high-energy scale  $M\sim{\cal O}($TeV$)$,  $\mu\ll M$, the coefficient of the $d=5$ operator necessarily  acquires an extra suppression in powers of $\mu/M$, while the fermionic $d=6$ operators keep its unsuppressed $1/M^2$ dependence. }
  
 As an example, a typical decoupling pattern goes qualitatively as follows:
  \be\label{cd5dec}
c^{d=5} =f(Y) \frac{ \mu } {  M^2} \,,
\ee  
\be\label{cd6dec}
c^{d=6} = g(Y) \, \frac{1}{|M|^2} \,,
\ee  
where $f$ and $g$ are some functions of the Yukawa couplings, implying a light neutrino mass matrix of the form
\be\label{mnudec}
m_\nu= -{ f(Y)v^2 \over {2}} \, \frac{ \mu} {  M^2}\,,
\ee 
while the effects of the $d=6$ operator, Eq.~(\ref{cd6dec}), are independent of $\mu$ and may be sizable for
generic Yukawa couplings, which may remain large and even ${\cal O}(1)$. Notice that such dependence has already been found above for the minimal version of the scalar-triplet mediated Seesaw  scenario (with $f\sim Y_\Delta$, $g\sim Y_\Delta^\dagger  Y_\Delta$, $\mu=\mu_\Delta$),  see Eqs.~(\ref{cd5scalar}) and (\ref{L4F}), suggesting 
the possibility $\mu_\Delta \ll M_\Delta$, $Y_{\Delta}\sim 1$. 
We call this universal pattern {\it direct lepton violation},  since the neutrino masses are proportional to the (small) lepton number violating quantity $\mu$, rather than inversely proportional to the large
lepton number violating heavy field mass.
\subsubsection*{ Multiple Seesaw models} 
Let us consider, as illustration and support of our general ansatz,  models existing in the literature and  based on extensions of the type-I 
Seesaw scenario with a low scale.
The examples considered below can be straightforwardly applied and extended to the type-III 
- triplet-fermion mediated- 
Seesaw scenarios.
We are thus interested in a class of models 
which, to lead to sufficiently suppressed neutrino masses and large $d=6$  
 operators, do not require any precise cancellations between the various (a 
priori independent) entries of the singlet neutrino 
mass  matrix
and/or of the Yukawa matrix\footnote{Consequently, we don't 
consider cases such as, for example, that in  Ref.~\cite{PilaftsisPRL}, based on 
the 
relation $Y_\nu Y_\nu^T=0$ and $(M_N)_{ij}=m_N\delta_{ij}$.}. The cases 
considered below just require that some of the entries of these mass matrices carry  
Majorana character and  are 
much smaller than other ones.
For simplicity,  only one left-handed neutrino and two 
singlet fermions will be included in the analysis $(\nu_L,N_1,N_2)$.
For instance, in the ``inverse Seesaw model"~\cite{GonzalezGarcia:1988rw}, the following texture is assumed\footnote{We acknowledge interesting discussions on this topic with S.~Antusch.}:
\begin{equation}
\left( \begin{array}{ccc}0 & m_{D_1} & 0 \\ 
m_{D_1} & 0 & M_{N_1} \\ 0 & M_{N_1} & \mu
\end{array} \right) \,,
\label{invsee}
\end{equation}
where $\mu$ is a small Majorana mass, $\mu \ll M_{N_1}$. All other entries in the matrix are of Dirac character: for $\mu=0$, assigning $L=1,-1,1$ to $\nu_L, N_1, N_2$ respectively, lepton 
number is indeed conserved by the Lagrangian and no Majorana mass results.
Expanding the eigenvalues of Eq.~(\ref{invsee}) in powers of $\mu/M_{N_1}$, a light eigenvalue is obtained:
\begin{equation}
m_\nu=\frac{m_{D_1}^2}{M_{N_1}}\frac{\mu}{M_{N_1}}\frac{M_{N_1}^2}{M_{N_1}^2
+m_{D_1}^2} +O(\mu^3)\,\,\longrightarrow_{(m_{D_1} \ll M_{N_1})}\,\frac{m_{D_1}^2}{M_{N_1}}\frac{\mu}{M_{N_1}}+O(\mu^3)\,,
\label{mnu1a}
\end{equation}
where higher order terms have been neglected.  As $m_{D_1}$ is a typical Dirac mass term, $m_{D�1}\sim{Y_1\,v/\sqrt{2}}$ with $Y_1$ a Yukawa coupling, 
Eq.~(\ref{mnu1a}) shows that the neutrino mass is suppressed by an extra factor $\mu/M_{N_1}$ with respect to the result for the minimal type-I Seesaw model, Eq.~(\ref{mnuI}), exactly as expected from the general argument above, see Eq.~(\ref{mnudec}):
\begin{equation}
m_\nu \sim \mu \frac{Y_1^2\,v^2}{M_1^2}\,.
\end{equation}
 The smallness of neutrino masses, and the 
argument of no fine-tuning, do not require tiny Yukawa couplings. 
For instance,  if the 
Yukawa coupling $Y_{\nu_1}$ is of order 
unity, i.e. $m_{D_1}=Y_{\nu_1} v\sim v$, and if $M_{N_1}\sim 1$ TeV, this 
requires $\mu/M_{N_1}\sim 10^{-12}$. Similarly, for $M_{N_1}\sim 1$ TeV, a rather  ``large'' Yukawa 
coupling 
of order $10^{-3}$ requires $\mu/M_{N_1}\sim 10^{-6}$.
 On the other hand, the $d=6$ operator coefficient is independent of $\mu$, as in Eq.~(\ref{cd6dec}), and low-energy effects associated to it  - such as non-unitary mixings in the weak currents and other signals- could be discovered in the near future. 
 
These results can be generalized to the case with the most general  matrices which, with large Yukawa couplings, still lead to vanishing
$\nu$ masses from extra small entries,
therefore leading to suppressed $d=5$ operator coefficients together with large $d=6$ operator coefficients. 
For instance, in the two $N$ plus one $\nu$ case above, the most general Majorana texture is the one in Eq.~(\ref{invsee}) with an additional non-zero value for the $22$ element. This can be justified for instance in the context of extended models (see e.g. Ref.~\cite{Dudas}).
 A non-zero $22$ entry has the interesting feature of being a source of lepton number violation without inducing by itself neutrino masses: for $\mu=0$ the determinant still vanishes leading to massless neutrinos. We will postpone the discussion of scenarios with a non-zero value for that entry to  Appendix C 
 , where a generalization to
the 3 left-handed plus 3 right-handed neutrino case can also be found. 
 Analogous extensions of fermion-triplet mediated type-III Seesaw models are straightforward. The interesting textures are just the same as in the type I (that is, singlet-fermion Seesaw) scenario.
 
In conclusion, irrespective of whether the Seesaw mechanism results from the exchange of heavy fermions or heavy scalars, to have large effects from $d=6$ operators requires first to lower the scale $M$ toward the TeV range and second a decoupling of the values of the $d=5$ and $d=6$ coefficients  along the pattern developed above, i.e. Eq.~(\ref{cd5dec}) and  Eq.~(\ref{cd6dec}).  This allows to account for the experimental values of neutrino masses without neither fine-tuning the Yukawa couplings nor assuming cancellations in combinations of them. For a Seesaw scale of ${\cal O}($TeV$)$, observable effects are then possible. The next Section - which deals with the phenomenological aspects of Seesaw models including bounds for any value of $M$  -  will focus on those effects.
\newpage
\section{Phenomenology of Seesaw models}
\label{pheno}
\subsection{Fermionic singlets}
\label{typeIpheno}
 The models where the heavy fields are SM singlets  are most difficult to test, as they lead  to fewer and rarer experimental 
 signals at low energies, even for low Seesaw scales. There exist, though, bounds on combinations of the Yukawa couplings which can be saturated for the type-I  inverse Seesaw and similar extensions, as well as for models with extra dimensions containing Kaluza-Klein replicas which are SM singlets~\cite{extradim}.
 
The bounds stem from important indirect signals which may be induced from the fact that the leptonic mass matrix appearing in the charged current is no longer unitary, see Section \ref{typeI}. This subject, as well as the determination of the corresponding bounds on $|N N^\dagger|_{\alpha \beta}$, has been studied at length recently \cite{uni}. In a nutshell, deviations from the values expected in a unitary analysis are constrained to be of order $1\%$ or smaller. Indeed, a global fit to the constraints resulting from $W$ decays, $Z$ decays, universality tests and rare lepton decays proved~\cite{uni} that the $NN^\dagger$ elements
 agree with those expected in the unitary case, within a precision better than a few percent, at the 90\% CL:
\begin{eqnarray}
|N N^\dagger | \approx
\begin{pmatrix}
 0.994\pm  0.005   & < 7.0 \cdot 10^{-5}  &   < 1.6 \cdot 10^{-2}\\
 < 7.0 \cdot 10^{-5}   &  0.995 \pm  0.005 &  < 1.0 \cdot 10^{-2}  \\
 < 1.6 \cdot 10^{-2}   &  < 1.0 \cdot 10^{-2}  &  0.995 \pm 0.005
\end{pmatrix} \, .
\label{nndag}
\end{eqnarray}
 The off-diagonal constraints in Eq.~(\ref{nndag}) result from the experimental bounds existing on the  radiative processes $\mu\rightarrow e \gamma$, $\tau\rightarrow e \gamma$ and $\tau\rightarrow \mu \gamma$, while the diagonal ones come from  the combined analysis of all other processes mentioned above.
    Using now the relation obtained in Eq.~(\ref{Nc6-N}) between the elements of the coefficient matrix $c^{d=6}$  and those of $NN^\dagger$, it follows that 
  \begin{eqnarray}
 \label{bounds-I}
\frac{v^2}{2}\,|c^{d=6}|_{\alpha\beta}\,=\,\frac{v^2}{2}\,|Y_N^\dagger\frac{1}{|M_N|^2}Y_N|_{\alpha\beta}\ \lesssim
\begin{pmatrix}
 10^{-2}  & 7.0 \cdot 10^{-5}  &    1.6 \cdot 10^{-2}\\
  7.0 \cdot 10^{-5}   &   10^{-2} & 1.0 \cdot 10^{-2}  \\
  1.6 \cdot 10^{-2}   &  1.0\cdot 10^{-2}  & 10^{-2}
\end{pmatrix} \, .
\label{limits}
\end{eqnarray}
When obtaining the numerical bounds in Eqs.~(\ref{nndag}) and
 (\ref{bounds-I}), the effective theory was used to compute
 $\mu\rightarrow e \gamma$ and $\tau\rightarrow \mu \gamma$, that is,
 the analysis was done in terms of $c^{d=5}$ and $c^{d=6}$. It is to
 be noticed that the computation of such one-loop transitions in the
 effective theory does not coincide exactly with that done in the full
 theory (i.e. type I Seesaw model), as higher dimension effective
 operators have to be taken into account in the matching between
 both. Numerically, the differences are of ${\cal O}(1)$ and
 irrelevant for the precision attempted here, though.  Furthermore,
 when computing these $l_1\rightarrow l_2 \gamma$ transitions - here
 and in the chapters to follow - we will not take into account the
 electromagnetic radiative corrections~\cite{Czarnecki:2001vf}, as
 their inclusion would correspond to a two-loop calculation and
 numerically they will not change the order of magnitude of the bounds
 obtained.
 
 Notice that the bounds above are valid for any value of $M_N$ and
  apply to any (type I) Seesaw theory. In consequence, they apply to
  the inverse Seesaw model considered above, in which $M_N$ could be
  near the TeV scale while the Yukawa couplings may be large, and
  signals could appear at the edge of the experimental limits above.
  New signals of CP-violation in neutrino oscillations, sensitive to
  the phases of $c^{d=6}$ may also be observable in future
  facilities~\cite{FernandezMartinez:2007ms}.
 
  As for direct detection of the heavy singlets in future accelerators in case $M_N\sim 1\,$TeV, several studies exist of the associated production of the heavy singlets and the Higgs particle, with difficult prospects for a positive signal~\cite{delAguila}.

   \subsection{Scalar triplets}
      \label{typeIIpheno}
     Using the experimental values for the renormalized parameters $\alpha$, $G_F$ and $M_Z$ as defined in the Z-scheme in Sect.~2.2, we will now consider deviations from the SM predictions for a set of observables.
\subsubsection{$\boldsymbol{M_W}$ and the $\boldsymbol{\rho}$ parameter}
The operator ${\cal L}_{\phi D}$ induces corrections to the predicted value of $M_W$ and to the $\rho$ parameter, 
 through the term $4\frac{|\mu_\Delta|^{2}}{|M_{\Delta}|^{4}}\left[\phi^{\dagger}D_{\mu}\phi\right]^{\dagger}\left[\phi^{\dagger}D_{\mu}\phi\right]$ in Eq.~(\ref{LfiD1}). When the $\rho$ parameter is extracted from data using only hadronic transitions\footnote{It is customary to extract the value of $\rho$ from a  global fit to data, including simultaneously hadronic and leptonic transitions; if the latter were considered in the analysis, further corrections would appear in $\delta \rho$, induced by $\delta {G_F}$ in Eq.~(\ref{G_F}).}, its predicted value is shifted  from the SM prediction by
  \begin{equation}
  \label{deltarho}
 \delta \rho_{\rm {had}}= -\frac{|\mu_\Delta|^{2}}{M_\Delta^{4}}\,\frac{\sqrt{2}}{G_F}\,,
 \end{equation}
 a result previously obtained in the literature~\cite{Gunion:1989we}.

 We find that the mass of the $W$ boson is also predicted to acquire a shift from both ${\cal L}_{\phi D}$ and ${\cal L}_{4F}$  in Eq.~(\ref{LfiD}), which is given by
 \begin{eqnarray}
 \label{deltaMW}
 \delta M_W^2 &=&-\frac{M_W^2}{2{M_W}^2 - M_Z^2} \left[
    \delta \rho_{had}\, M_W^2\,-\, \frac{\delta {G_F}}{G_F}
  (M_W^2-M_Z^2)\right]\\
 &=&
  -\frac{M_W^2}{2{M_W}^2 - M_Z^2} \left[
  \frac{|\mu_\Delta|^{2}}{M_\Delta^{4}}\,
  \frac{M_W^2}{G_F\sqrt{2}} \,-\, 
  \frac{M_W^2-M_Z^2}{\sqrt{2}G_FM_{\Delta}^{2}} {Y_{\Delta}}_{e\mu }{Y_{\Delta}^{\dagger}}_{e\mu }\right]\ .\nonumber
  \end{eqnarray} 
  
  In this equation, 
 $M_W$  is to be identified with the SM prediction for the $W$-boson mass in the $Z$-scheme, 
 \begin{eqnarray}
  \label{MWSM}
  (M_W^{SM})^2= 
  \frac{M_Z^2}{2}\,\left(1+
  \sqrt{
  1- \frac{4\pi \alpha }{\sqrt{2}G_F M_Z^2}}\right)
  \,\frac{1}{(1-\Delta r)}\ ,
  \end{eqnarray}
  where $\Delta r$ accounts for the dominant SM one-loop radiative corrections\footnote{While these corrections are important when compared to the total value of $M_W$, they can be dropped in Eq.~(\ref{deltaMW}), as we work at first order in all corrections. 
  }~\cite{PDG}, and $G_F$ is extracted from muon decay, see Eqs.~(\ref{G_F}) 
  and (\ref{GFMscalar}).

The very precise experimental determination  of the $W$ boson mass allows to set stringent bounds on both terms in Eq.~(\ref{deltaMW}), barring extreme cancellations between both.
From the difference between the experimental value and the SM prediction of $M_W$ obtained in the Z-scheme  ($M_W^{SM}=(80.4887\pm 0.0515)$ GeV) we obtain:
\begin{eqnarray}
|{Y_{\Delta}}_{\mu e}|^4
=\left(0.00023\pm {0.00109}\right)\left(\frac{M_{\Delta}}{1\;\rm{TeV}}\right)^{-4} \,,
\label{MWbound}
\end{eqnarray} 
and
\begin{equation}
- v^2\frac{|{\mu_\Delta}|^{2}}{{M_\Delta}^{4}}=0.0001368\pm {0.00032},\ \  {\rm or}
\end{equation}
\begin{equation}
\frac{|{\mu_\Delta}|}{M_{\Delta}^{2}}< 8.7\times 10^{-2}\, {\rm TeV}^{-1}\ .
\label{YDeltaemu}
\end{equation}
Notice however that the hadronic data on the $\rho$  parameter allow to independently constraint $\mu_\Delta/M_\Delta^2$. As an estimate, taking at face value the average experimental value of the $\rho$ parameter ($\rho=1.0002\pm^{0.0007}_{0.0004}$) as if it were indeed dominated by the hadronic contributions - which do not depend on the leptonic Yukawa couplings -, it would follow that
\begin{equation}
- v^2\frac{|{\mu_\Delta}|^{2}}{{M_\Delta}^{4}}< 0.0001\pm^{0.00035}_{0.0002}\ .
\end{equation}
The neutrino masses in Eq.~(\ref{mnutscal}) are given by the square
root of this ratio multiplied by $Y_\Delta$. For instance, $m_\nu
\sim1$eV and $Y_\Delta\sim {\cal O} (1)$ requires
${\mu_\Delta}/M_{\Delta}^{2}\sim10^{-11}$eV, well below the bound in
Eq.~(\ref{YDeltaemu}), while the lower limit of the bound can be
saturated for values of $Y_\Delta\sim10^{-7}$.
 
 \subsubsection{$\boldsymbol{\mu} \rightarrow \boldsymbol{eee}$ and $\boldsymbol{\tau} \rightarrow 3 \boldsymbol{l}$ decays}
 
$\delta {{\cal L}_{4F}}$ in Eqs.~(\ref{L4F}) and (\ref{L4Ffamiliar}) induces exotic four-lepton couplings contributing to lepton-flavor violating processes.
   Notice that this operator does not depend on the scale $\mu_\Delta$, so that the discussion is independent of it.
Besides its impact on the determination of $G_F$ from muon decay,  Eqs.~(\ref{G_F}) and (\ref{GFMscalar}), 
  it modifies the branching ratios for rare leptonic decays. 
 The constraints implied by the present experimental bounds on these processes 
  have been studied in models with a scalar triplet in  Refs.\cite{Glas}-\cite{Gunion2}.
 
Important decays are $\mu^-\rightarrow e^{+}e^{-}e^{-}$ and $\tau \rightarrow 3l$, considered in the full theory with a scalar triplet in Refs.~\cite{Barg,Pal} and from the generic leptonic $d=6$ effective operator  $\delta {{\cal L}_{4F}}$ in 
Refs.\cite{Cahn:1980kv}-\cite{Cirigliano:2006su}.  In terms of the 
 coefficient $c^{4F}_{\alpha \beta \gamma \delta}$ of this leptonic operator in Eq.~(\ref{L4F}),
 \begin{equation}
 c^{4F}_{\alpha \beta \gamma \delta} \equiv {1\over  {M}_{\Delta}^{2}}
Y_{\Delta_{\alpha \beta }}\,Y_{\Delta_{\gamma\delta}}^\dagger\,,
 \end{equation}
 we obtain
\begin{equation}
\Gamma(\mu^-\rightarrow e^+ e^- e^-)=
{m_{\mu}^5 \over 192 \pi^3} {\left|c^{4F}_{\mu e ee}\right|^2  }=
{m_{\mu}^5 \over 192 \pi^3} {1\over {M}_{\Delta}^{4}}
|Y_{\Delta_{\mu e }}|^2 |Y_{\Delta_{ee}}|^2
\end{equation}
which gives
\bea
\textrm{Br}(\mu^- \rightarrow e^+ e^- e^-)\simeq
\frac{\Gamma(\mu^- \rightarrow e^+ e^- e^-)}
{\Gamma(\mu^- \rightarrow e^- \nu_\mu \overline{\nu}_e)}
={\left|c^{4F}_{\mu e ee}\right|^2 \over  G_F^2}
= {1\over  {M}_{\Delta}^{4}G_F^2}|Y_{\Delta_{e\mu }}|^2 |Y_{\Delta_{ee}}|^{2}.
\eea
Similarly we obtain
\begin{equation}
\Gamma(\tau^-\rightarrow l_i^+ l_j^- l_j^-)={m_\tau^5 \over 192 \pi^3} {\left|c^{4F}_{\tau ijj}\right| }
={m_\tau^5 \over 192 \pi^3} {1\over  {M}_{\Delta}^{4}}
|Y_{\Delta_{\tau i} }|^2 |Y_{\Delta_{jj}}|^2
\end{equation}
for any $i$ and $j$, while
\begin{equation}
\Gamma(\tau^-\rightarrow l_i^+ l_j^- l_k^-)={m_\tau^5 \over 96 \pi^3} {\left|c^{4F}_{\tau ijk}\right| }
={m_\tau^5 \over 96 \pi^3} {1\over {M}_{\Delta}^{4}}
|Y_{\Delta_{\tau i} }|^2 |Y_{\Delta_{jk}}|^2
\end{equation}
for any $i,j,k$ with $j\neq k$.
Using all experimental branching ratios or upper limits on branching ratios as given in Ref.~\cite{PDG}, the corresponding bounds on the Yukawa couplings are given in Table~\ref{boundstree} .
  \begin{table}[h]
\begin{center}
\begin{tabular}{|c|c|c|}
\hline
Process 
& Constraint on & $\textrm{Bound} \left(\times\left(\frac{M_{\Delta}}{1\, {\rm TeV}}\right)^{2}\right)$ \\
 
 \hline
 
 $M_W$ 
& 
$|{Y_\Delta}_{\mu e}|^2$
& $<7.3\times 10^{-2}$   \rule[-8pt]{0pt}{22pt}\\
\hline
 
 $\mu^{-}\rightarrow e^{+}e^{-}e^{-}$ 
& 
$|{Y_\Delta}_{\mu e}||{Y_\Delta}_{ee}|$
& $<1.2\times 10^{-5}$   \rule[-8pt]{0pt}{22pt}\\
\hline
$\tau^{-}\rightarrow e^{+}e^{-}e^{-}$ 
& 
$|{Y_\Delta}_{\tau e}||{Y_\Delta}_{ee}|$
&  $<1.3\times10^{-2}$   \rule[-8pt]{0pt}{22pt}\\
 
\hline
 $\tau^{-}\rightarrow \mu^{+}\mu^{-}\mu^{-}$ 
& 
$|{Y_\Delta}_{\tau \mu}||{Y_\Delta}_{\mu \mu}|$
& $<1.2\times10^{-2}$    \rule[-8pt]{0pt}{22pt}\\
 
   \hline
$\tau^{-}\rightarrow \mu^{+}e^{-}e^{-}$ 
& 
$|{Y_\Delta}_{\tau \mu }||{Y_\Delta}_{ee}|$
& $<9.3\times10^{-3}$ \rule[-8pt]{0pt}{22pt} \\
 
  \hline
$\tau^{-}\rightarrow e^{+}\mu^{-}\mu^{-}$ 
& 
$|{Y_\Delta}_{\tau e}||{Y_\Delta}_{\mu \mu}|$
& $<1.0\times10^{-2}$  \rule[-8pt]{0pt}{22pt}\\
 
\hline
 $\tau^{-}\rightarrow \mu^{+} \mu^{-} e^{-}$ 
&
$|{Y_\Delta}_{\tau \mu }||{Y_\Delta}_{\mu e}|$
& $<1.8\times10^{-2}$    \rule[-8pt]{0pt}{22pt}\\

\hline
 
 $\tau^-\rightarrow e^{+} e^{-} \mu^{-} $ 
& 
 $|{Y_\Delta}_{\tau e}||{Y_\Delta}_{\mu e}|$
 & $<1.7\times10^{-2}$     \rule[-8pt]{0pt}{22pt}\\

\hline
\hline
 
$\mu\rightarrow e \gamma$ 
&$|\Sigma_{l=e,\mu,\tau}{Y_\Delta}_{l\mu}^{\dagger}{Y_\Delta}_{e l}|$ & $<4.7\times 10^{-3}$   \rule[-8pt]{0pt}{22pt}\\
 
\hline
  $ \tau\rightarrow e \gamma$ 
&  $|\Sigma_{l=e,\mu,\tau}{Y_\Delta}_{l\tau }^{\dagger}{Y_\Delta}_{e l}|$ & $<1.05$   \rule[-8pt]{0pt}{22pt} \\
 
  \hline
 $ \tau\rightarrow \mu \gamma$ 
 & $|\Sigma_{l=e,\mu,\tau}{Y_\Delta}_{l\tau }^{\dagger}{Y_\Delta}_{ \mu l}|$ & $<8.4\times 10^{-1}$  \rule[-8pt]{0pt}{22pt} \\
 
 \hline
\end{tabular}
\end{center}
\caption{Bounds on ${Y_\Delta}_{ij}$ from $M_W$, Eq.~(\ref{MWbound}), from tree level $\ell_1^-\rightarrow \ell_2^-\ell_3^+\ell_4^-$ decays and from one loop $l_1 \rightarrow l_2 \gamma$ processes.}
\label{boundstree}
\end{table}
For Yukawa couplings of order unity, the present non-observation of those LFV transitions, in particular of the most stringent one, $\mu \rightarrow eee$, implies a lower bound on the scalar triplet Seesaw scale,
 \be
 M_\Delta \ge 294 \ \text{TeV}\,\, , \qquad \text{for}\,\, Y_\Delta\sim {\cal O}(1) \,.
 \ee
\subsubsection{Complete Lagrangian  and  $\boldsymbol{l_1}\rightarrow \boldsymbol{l_2 \gamma}$}
 It is useful to consider also the bounds which arise from 
the radiative decays  $\ell_1\rightarrow\ell_2\gamma$, 
although these processes cannot be obtained completely from the $d=6$  operators because they are one-loop processes. Consider then instead the full high-energy  Lagrangian for the scalar triplet in Eq.~(\ref{ScalarLcomponents}), expanded into charge components: 
\begin{eqnarray}
\!\!\mathcal{L}_\Delta\! \!\!& = &\!\! \!\!D_{\mu}{\Delta^0}^* D^{\mu} \Delta^0+
D_{\mu}{\Delta^+} D^{\mu} \Delta^-+D_{\mu}\Delta^{++} D^{\mu} \Delta^{--}\nonumber\\
&+&
\left\{   (\overline{{l_L}^c} Y_\Delta \nu_L) \Delta^+ 
+ (\overline{{\nu_L}^c} Y_\Delta e_L )\Delta^+ +\sqrt{2} (\overline{{l_L}^c} Y_\Delta l_L )\Delta^{++}-\sqrt{2} (\overline{{\nu_L}^c} Y_\Delta \nu_L )\Delta^{0}
 + \text{h.c.}\right\}\nonumber\\
&+&\left\{\mu_\Delta\left(2\phi^0\phi^+\Delta^-+\sqrt{2}\phi^0\phi^0{\Delta^0}^*-\sqrt{2}\phi^+\phi^0\Delta^{--}\right)+\text{h.c.}\right\}\nonumber\\
\! \!\!&-& \!\!\!\!\! {M_\Delta}^{2}\left(\Delta^{0\,*}\Delta^0+\Delta^{-}\Delta^++\Delta^{--}\Delta^{++}\right)-{\lambda_2\over 2}\left(\Delta^{0\,*}\Delta^0+\Delta^{-}\Delta^++\Delta^{--}\Delta^{++}\right)^2\nonumber\\
&-&\lambda_3(\phi^-\phi^++\phi^{0*}\phi^0)\left(\Delta^{0\,*}\Delta^0+\Delta^{-}\Delta^++\Delta^{--}\Delta^{++}\right) -
\lambda_4\Big[{1\over 2} 
(\Delta^{0\,*}\Delta^0)^2+{1\over 2}(\Delta^{++}\Delta^{--})^2\nonumber\\&+& \Delta^{++}\Delta^{--}(\Delta^+\Delta^--\Delta^{0\,*}\Delta^0) +\Delta^+\Delta^-\Delta^{0\,*}\Delta^0 -\left\{\Delta^{++}\Delta^{-}\Delta^-\Delta^0+\text{h.c.}\right\}\Big] \label{ScalarLcomponents}\\
&-&\lambda_5\left[(\Delta^{++}\Delta^{--}-\Delta^{0\,*}\Delta^0)(\phi^+\phi^--\phi^{0\,*}\phi^0)-\sqrt{2}\left\{\phi^-\phi^0(\Delta^{++}\Delta^--\Delta^{0\,*}\Delta^+)+\text{h.c.}\right\}\right]\nonumber\,.
\end{eqnarray} 
\subsubsection*{$\boldsymbol{\mu}\rightarrow \boldsymbol{e\gamma}$, $\boldsymbol{\tau}\rightarrow \boldsymbol{e\gamma}$, $\boldsymbol{\tau}\rightarrow \boldsymbol{\mu \gamma}$}
 Radiative processes are due to the exchange between lepton fields of  both the $\Delta^{++}$ and $\Delta^{+}$ fields, as given in Eq.~(\ref{ScalarLcomponents}), and the branching ratios read~\cite{Bernabeu1,Bilenky,Mohapatra}: 
 \begin{eqnarray}
\textrm{Br}(l_{1}\rightarrow l_{2}\gamma)={\alpha \over 48 \pi} {25\over 16} \frac{\left|\underset{l}{\Sigma}{Y_\Delta}_{ll_{1}}^{\dagger}{Y_\Delta}_{l_{2}l}\right|^{2}}{G_{F}^{2}{M}_{\Delta}^{4}}\hbox{Br}(l_1\rightarrow e \nu_1 \bar{\nu}_e)\, .
\end{eqnarray}
The corresponding bounds are also given in Table~\ref{boundstree}.
Combining all bounds of this Table,
we have obtained new bounds for combinations of Yukawa parameters, not considered previously in the literature and gathered in Table 5. They show that, for low values of the Seesaw scale, the Yukawa couplings may be of ${\cal O}(1)$, while they should be sizeably smaller by up to 2 orders of magnitude for some specific flavours, for an ${\cal O}$(TeV) Seesaw scale.

\begin{table}
\begin{center}
\begin{tabular}{|c|c|c|}
\hline
\multicolumn{3}{|c|}{Combined bounds} \\
\hline
Process & Yukawa & $\textrm{Bound} \left(\times\left(\frac{M_{\Delta}}{1\, {\rm TeV}}\right)^{4}\right)$\\
 
 \hline
 
$\mu\rightarrow e \gamma$ & $\left|{Y_\Delta}_{\mu\mu}^{\dagger}{Y_\Delta}_{\mu e}+{Y_\Delta}_{\tau\mu}^{\dagger}{Y_\Delta}_{\tau e}\right|$ & $<4.7\times10^{-3}$ \rule[-8pt]{0pt}{22pt} \\
 
\hline
  $ \tau\rightarrow e \gamma$ &  $\left|{Y_\Delta}_{\tau\tau}^{\dagger}{Y_\Delta}_{\tau e }\right|$ & $<1.05$ \rule[-8pt]{0pt}{22pt}  \\
 
  \hline
 $ \tau\rightarrow \mu \gamma$ & $\left|{Y_\Delta}_{\tau\tau}^{\dagger}{Y_\Delta}_{\tau \mu}\right|$ & $<8.4\times10^{-1}$ \rule[-8pt]{0pt}{22pt} \\
 
 \hline
\end{tabular}
\end{center}
\label{boundscom}
\caption{Bounds on combinations of ${Y_\Delta}_{ij}$. }
\end{table}

\subsubsection{Other constraints}
\noindent
There are also other bounds which arise from other processes, from  
Bhabha scattering \cite{Barg,Gunion1,Mohapatra} (leading to the bound  
$|Y_{\Delta_{ee}}| < 1.0 \cdot (M_\Delta/1$ TeV))
, from muonium to antimuonium
conversion \cite{Barg,Gunion2} (leading to the bound $|Y_{\Delta_ 
{ee}}| |Y_{\Delta_{\mu \mu}}| <0.1 \cdot (M_\Delta/1$ TeV$)^2$), and  
from the anomalous magnetic moment of the muon. The
latter constraint
comes from the fact that the
doubly charged scalar  $\Delta^{\pm\pm}$ as well as the simply charged  $\Delta^{\pm}$ 
induce a shift in the anomalous magnetic moment of
the muon \cite{Gunion1,Gunion2}\,,
\begin{equation}
\delta (a_\mu)=-\frac{m^2_\mu}{3 \pi M^2_\Delta} \sum_{j=e,\mu'\tau}| 
Y_{\Delta_{\mu j}}|^2\,,\quad
\delta (a_\mu)=-\frac{m^2_\mu}{24 \pi M^2_\Delta} \sum_{j=e,\mu'\tau}| 
Y_{\Delta_{\mu j}}|^2\, ,
\end{equation}
respectively. This contribution
has opposite sign with respect to the observed deviation. Taking for  
instance $\delta (a_\mu)< 20 \times 10^{-10}$, we get $\sum_{j=e, 
\mu'\tau}|Y_{\Delta_{\mu j}}|^2< 1.9 \cdot (M_\Delta/1$ TeV$)^2$.
\subsubsection{Collider signatures of scalar triplets}
  
Scalar triplet Seesaw opens the possibility of observing new signals at present and/or future facilities.
For instance, for $Y_\Delta\sim {\cal O}(1)$, a positive observation of 
$\mu \rightarrow e \gamma$ by the MEG experiment \cite{MEG} (which aims to achieve $10^{-13}$ sensitivity for the branching ratio)
would require 
 \be
 15\,\, {\rm TeV}\ < M_\Delta < 50\,\, {\rm TeV} \,, \label{rangesca}
 \ee
while $Y_\Delta\sim {\cal O}(10^{-2})$ would require 
$0.15 \,\, {\rm TeV}\ < M_\Delta <  0.5\,\, {\rm TeV}$.
If $M_\Delta$ turns out to be as low as ${\cal O}($TeV$)$,  the  non-vanishing electroweak charge of the $\overrightarrow\Delta$ field  offers the possibility  of  clean  signals in hadronic accelerators (Tevatron, LHC). The  production (or associated production) of $\Delta^{++}$ and $\Delta^{--}$ particles,   and  their subsequent decay in pairs of same-sign leptons, would constitute   striking signals, free from SM backgrounds~\cite{Akeroyd,Accomando}.
 A lower bound on the mass of  $\Delta^{\pm\pm}$ of the order of $136$ GeV has been obtained at CDF \cite{Acosta}.
  Assuming that  a boson with those characteristics is indeed observed in an accelerator, one still needs to ascertain whether a scalar-mediated Seesaw mechanism 
  is indeed at work. For that, and as a first step, it is necessary  to measure
  and 
 disentangle the Yukawa couplings appearing in Tables \ref{boundstree} and 5.
  In order to extract values for the individual ${Y_\Delta}_{ij}$, it would be necessary to observe in addition at least three lepton flavor violating processes.

  The first term in the Lagrangian in Eq. (\ref{ScalarL}) generates tree-level vertices $\Delta^{\pm\pm} W^\pm W^\pm$ and $\Delta^{\pm} W^\mp Z$, which would be detected by observing for instance   $\Delta^{++}\to W^+W^+$,  $W^+W^+\to \Delta^{++}$ , $Z^*\to\Delta^{+}W^-$, or  $\Delta^{+}\to Z W^+$. The analysis of some of these processes has already  been covered in \cite{Raidal}, \cite{Hektor},  for the LHC. 
Once produced by Drell-Yann processes ($\bar q q\to \Delta^{--} \Delta^{++}$), with production cross-section given in Refs.\cite{Akeroyd,Accomando}, the $\Delta^{++}$ ($\Delta^{--}$) can 
 decay into pairs of $W$s  of the same charge ($\Delta^{\mp\mp}\to W^\mp W^\mp$), for which the decay rate is proportional to  $v^2\frac {M_\Delta^3}{M_W^2}\frac{|\mu_\Delta|^2}{M_\Delta^4}$, or into leptons $l_i,l_j$ of the same charge, with  a decay rate  proportional to $M_\Delta|{Y_\Delta}_{ij}|^2$. Finally, the $\Delta^{++}$ ($\Delta^{--}$) particle can also decay into a charged Higgs and an off-shell $W$ gauge boson, as in $\Delta^{--}\to \phi^- W^{*-}$. The decay rate of the latter process is suppressed when compared to the previous ones, unless the  $\lambda_5$ coupling in Eq. (\ref{ScalarL}) takes an unnaturally large value~\cite{Raidal},~\cite{Chun2003}.
Due to the constraint obtained in Eq. (\ref{YDeltaemu}), the process $\Delta^{--}\to W^-W^-$ will be suppressed and the only  relevant channel in our scenario will be  $\Delta^{\pm\pm}\to l^\pm l^\pm$, which will be background-free. 
 The related branching ratio will give access to $|{Y_\Delta}_{ij}|$, which is directly related to neutrino mass matrix elements up to the global factor $\frac{|\mu_\Delta|}{M_\Delta^2}$, i.e. to the effective theory coefficient $
c^{d=5}$ in Eq. (\ref{cd5scalar}).
 
Other interesting signals can be also searched for in accelerators.
  ${\cal L}_{6\phi }$ (Eq. (\ref{L6fi})) and the first term  of ${\cal L}_{\phi D}$ in Eq.~(\ref{LfiD1}), besides modifying the Higgs potential and renormalizing the scalar kinetic energy term, induce new couplings: $HWW$, $HZZ$, $HHWW$, $HHZZ$, $H^3$ and $H^4$ -where $H$ stands for the physical Higgs-. Consequently, the Higgs production cross sections at the future facilities ILC and CLIC~\cite{Barger2003} get corrections. Nevertheless, the 
 strong limit in Eq.~(\ref{YDeltaemu}) precludes observable effects, except maybe from  ${\cal L}_{6\phi }$ for very large values of $\lambda_3$ and/or  $\lambda_5$  ~\cite{Barger2003}.

  Similarly, ${\cal L}_{\phi D}$  
also affects Higgs physics. Its impact  on the Higgs decay branching ratios  has been analyzed~\cite{Kanemura}, although again the bound from the $\rho$ parameter discussed above excludes observation in the planned future facilities such as ILC.
 
  
\subsection{Fermionic triplets} 
\label{sectfermtrip}
We have argued that non-unitary flavour-changing matrices are to be
expected in this case for the couplings of light leptons to the W and
Z gauge boson, see Eqs.~(\ref{JCC_fermtrip})-(\ref{JNCe_fermtrip}).
The putative departures from unitarity can be re-expressed directly in
terms of the $d=6$ operator coefficients, that is to say in terms of the Yukawa couplings, see
Eq.~(\ref{Nc6-ft}). Specifically, notice that\footnote{ Again, the absolute-value bars in Eq.~(\ref{aa}) can be dropped when choosing the appropiate basis in flavour space, see the discussion in Sect.~\ref{fermtriptheory}.}
\begin{eqnarray}
&\,& |NN^\dagger-1|= |\epsilon^\Sigma|\,,\label{aa}\\
&\,& (N^\dagger N)^{-1}= 
U_\nu^\dagger\,(1\,-\,\epsilon^\Sigma)\,U_\nu 
\sim 1\,-\,U_{PMNS}^\dagger\,\epsilon^\Sigma\, U_{PMNS}\,.
\label{NNdagminus1}
\end{eqnarray}
For values of $M_\Sigma$ close to the electroweak scale, the
deviations of these quantities from their standard values  can be at the edge
of the present experimental bounds on non-unitarity.  Taking into
account the shift induced on $G_F$ as extracted from muon decay,
Eq.~(\ref{G_F_fermtrip}) for the effective theory, we proceed to
compute below the departures predicted on leptonic processes in the
effective and full theories.
 As we will see, all transitions
considered below result in constraints on the elements of the
$NN^\dagger$ matrix - and thus on the $d=6$ operator coefficients -,
analogously to the situation for fermionic singlet Seesaw
theories~\cite{uni}, see for instance Eq.~(\ref{bounds-I}). Indeed, even
if we could have expected that $Z$-mediated processes are sensitive also to
$U_{PMNS}$ through Eq.~(\ref{NNdagminus1}), this is not the case, as
we will show in Sect.~\ref{Z-decays}.
\subsubsection{$\boldsymbol{W}$ decays}
\label{W-decays}
The non-unitary mixing matrix $N$ appearing now in the charged weak
couplings, Eq.~(\ref{JCC_fermtrip}), results in a leptonic $W$ decay
width of the form
\begin{eqnarray}
\label{Eq:Wdecay1}
\Gamma (W \rightarrow l_\alpha \nu_\alpha ) =
\sum_i\Gamma (W \rightarrow l_\alpha \nu_i ) =
\frac{G_F^{SM} M_W^3}{6 \sqrt{2} \pi} (N N^\dagger)_{\alpha\alpha} \, .
\end{eqnarray}
Using the value of $G_F$ 
extracted from the decay $\mu \rightarrow \nu_\mu e \bar{\nu}_e$, as
given in Eq.~(\ref{G_F_fermtrip}), the following combinations can be
defined:
\begin{eqnarray}
\frac{(NN^\dagger)_{\alpha\alpha}}
{\sqrt{(NN^\dagger)_{ee}(NN^\dagger)_{\mu\mu}}}
= \frac{\Gamma (W \rightarrow \ell_\alpha \nu_\alpha ) \, 
   6 \sqrt{2} \pi}{G_F M_W^3}
\equiv f_\alpha \, .
\end{eqnarray}
With the experimental values of the $W$ decay widths and mass from
Ref.~\cite{PDG} and $G_F = (1.16637 \pm 0.00001) \times 10^{-5}$, the
parameters $f_\alpha$ take the values:
\begin{eqnarray}
\nonumber
f_e &=& 1.000\pm  0.024\, ,\\
\nonumber
f_\mu &=& 0.986 \pm  0.028\, ,\\
f_\tau &=& 1.002 \pm 0.032 \, .
\end{eqnarray}
\subsubsection{ Invisible $\boldsymbol{Z}$ decay}
\label{Z-decays}
The modified neutral weak couplings in Eqs.~(\ref{JNCnu_fermtrip}) and
(\ref{JNCe_fermtrip}) lead to
\begin{eqnarray}
\label{iag}
\Gamma (Z \rightarrow \mbox{invisible} ) =  
\sum_{i,j}\Gamma (Z \rightarrow \bar{\nu}_i \nu_j ) = 
\frac{G_F^{SM} M_Z^3}{12 \sqrt{2} \pi}\ (1 + \rho_t)
\,\sum_{i,j} |[(N^\dagger N)^{-1}]_{ij}|^2\,,
\end{eqnarray}
where $\rho_t \approx 0.008$~\cite{PDG} takes into account radiative
corrections mainly stemming from loops including the top quark. As the
dominant radiative corrections do not involve leptons, the
dependence on the mixing matrix in Eq.~(\ref{iag}) appears as a global
factor to an excellent approximation.  Using the data provided in
Ref.~\cite{PDG} and the following approximation valid at first order in
$\epsilon^\Sigma$
\be
\label{myapprox}
\sum_{i,j} |[(N^\dagger N)^{-1}]_{ij}|^2 = 
Tr (1-2\,\epsilon^\Sigma)= 
9-2\sum_\alpha (NN^\dagger)_{\alpha\alpha}\, , 
\ee
the following constraint is then obtained:
\begin{eqnarray}
\frac{9-2\sum_\alpha (NN^\dagger)_{\alpha\alpha}}
{\sqrt{(NN^\dagger)_{ee}(NN^\dagger)_{\mu\mu}}}
= \frac{12 \sqrt{2} \pi \,\Gamma (Z \rightarrow \mbox{invisible} )}
{G_F M_Z^3(1+\rho_t)}
= 2.984 \pm 0.009 \,.
\end{eqnarray}
As it is well known, this number should correspond to the number of
active neutrinos at LEP. Its $2\sigma$ departure from the value of $3$
is not (yet) significant enough to be interpreted as a signal of new
physics.
\subsubsection{Universality tests}
\label{univ-tests}
The existing constraints on the universality of weak interactions can
be turned into bounds on non-unitarity if the weak couplings are
indeed universal, as it is the case in Seesaw models. The results of
our analysis, always at order $\epsilon^\Sigma$, are displayed in
Table~\ref{unitests},  where the bounds have been extracted from Ref.~\cite{Pich:2005mk}.  
\begin{table}[!h]
\centering 
\begin{eqnarray*}
\begin{array}{|c|c|c|} 
\hline 
\displaystyle \mbox{Constraints on}  \vphantom{\frac{a}{a}}
&
\mbox{Process}
&
\mbox{Bound}
\\
\hline\hline
\displaystyle\frac{(N N^\dagger)_{\mu\mu}}{(N N^\dagger)_{ee}}
&
\displaystyle\frac{\Gamma (W \rightarrow \mu \bar{\nu}_\mu )}{
      \Gamma (W \rightarrow e \bar{\nu}_e )}
&
0.997 \pm 0.010  \rule[-14pt]{0pt}{34pt}
\\
\hline
\displaystyle\frac{(N N^\dagger)_{\tau\tau}}{(N N^\dagger)_{ee}}
&
\displaystyle\frac{\Gamma (W \rightarrow \tau \bar{\nu}_\tau )}{
      \Gamma (W \rightarrow e \bar{\nu}_e )}
&
1.034 \pm 0.0014  \rule[-14pt]{0pt}{34pt}
\\
\hline
\displaystyle\frac{(N N^\dagger)_{\mu\mu}}{(N N^\dagger)_{ee}} 
&
\displaystyle\frac{\Gamma (\pi \rightarrow \mu \bar{\nu}_\mu )}{
      \Gamma (\pi \rightarrow e \bar{\nu}_e )}
&
1.0017 \pm 0.0015  \rule[-14pt]{0pt}{34pt}
\\
\hline
\displaystyle\frac{(N N^\dagger)_{\tau\tau}}{(N N^\dagger)_{\mu\mu}}
&
\displaystyle\frac{\Gamma (\tau \rightarrow \pi \bar{\nu}_\tau)}{
      \Gamma (\pi \rightarrow \mu \bar{\nu}_\mu )}
&
0.9999 \pm 0.0036  \rule[-14pt]{0pt}{34pt}
\\
\hline
\displaystyle\frac{(NN^\dagger)_{\mu\mu}} 
{(NN^\dagger)_{ee}}
&
\displaystyle\frac{\Gamma (\tau \rightarrow \nu_\tau \mu \bar{\nu}_\mu )}{
      \Gamma (\tau \rightarrow \nu_\tau e \bar{\nu}_e )}
&
0.9999 \pm 0.0020  \rule[-14pt]{0pt}{34pt}
\\
\hline 
\displaystyle\frac{(NN^\dagger)_{\tau\tau}} 
{(NN^\dagger)_{\mu\mu}}
&
\displaystyle\frac{\Gamma (\tau \rightarrow \nu_\tau  e \bar{\nu}_e)}{
      \Gamma (\mu \rightarrow \nu_\mu e \bar{\nu}_e )}
&
1.0004 \pm 0.0023     \rule[-14pt]{0pt}{34pt}
\\
\hline
\displaystyle\frac{(NN^\dagger)_{\tau\tau}}
{(NN^\dagger)_{ee}}
&
\displaystyle\frac{\Gamma (\tau \rightarrow \nu_\tau \mu \bar{\nu}_\mu )}{
      \Gamma (\mu \rightarrow \nu_\mu e \bar{\nu}_e )}
&
1.0002 \pm 0.0022  \rule[-14pt]{0pt}{34pt}
\\
\hline
\end{array} 
\end{eqnarray*} 
\caption{\it Constraints on $(N N^\dagger)_{\alpha\alpha}$ from a selection of processes.}
\label{unitests}  
\end{table}
For the leptonic decays, the following expression has been used (for $\alpha\ne\beta$):
\be
\Gamma (l_\alpha \rightarrow  \nu_\alpha l_\beta \overline{\nu}_\beta ) =\frac{{G_F^{SM}}^2 m_\alpha^5}{192 \pi^3}
(NN^\dagger)_{\alpha\alpha}(NN^\dagger)_{\beta\beta} \, . \ee
Charged pion decays to a lepton pair are also considered in that Table.
%
\subsubsection{$\boldsymbol{Z}$ decays into charged leptons}
\label{Z-decays-charged}
While the processes analyzed in the previous Sections permit to put
bounds on the diagonal elements of $(NN^\dagger)$, as in the case of
the fermionic singlets, the additional presence of flavour changing effects  in the
coupling of charged fermions to the $Z$ boson allows to constrain the
off-diagonal elements of $(NN^\dagger)$ with tree-level processes, at variance
 with the fermionic singlet case. 
The leptonic width of the $Z$ gauge boson is given by
\begin{eqnarray}
\label{Zcharged}
\Gamma (Z \rightarrow l_\alpha \overline{l}_\alpha ) = 
\frac{G_F^{SM} M_Z^3}{3 \sqrt{2} \pi}
( | \sin^2\theta_W |^2 + 
 | \sin^2\theta_W - \frac{1}{2} [(NN^\dagger)^2]_{\alpha\alpha} |^2 ) \, ,
\end{eqnarray}
where the first (second) term in the parenthesis is the contribution
of right-handed (left-handed) leptons. For $\alpha\ne\beta$ it follows
that:
\begin{eqnarray}
\label{Zcahrged}
\Gamma (Z \rightarrow l_\alpha \overline{l}_\beta ) = 
\frac{G_F^{SM} M_Z^3}{3 \sqrt{2} \pi}
\frac{1}{4} |[(NN^\dagger)^2]_{\alpha\beta} |^2 \, .
\end{eqnarray}
It is now possible to obtain the branching ratios at leading
order in $\epsilon^\Sigma$:
\bea
\textrm{Br}(Z \rightarrow l_\alpha \overline{l}_\beta)&=&
\frac{\Gamma (Z \rightarrow l_\alpha \overline{l}_\beta )}
{\Gamma (Z \rightarrow l_\gamma \overline{l}_\gamma )}
\textrm{Br}(Z \rightarrow l_\gamma \overline{l}_\gamma)=\\
&=&\frac{|(NN^\dagger)_{\alpha\beta} |^2}
{2\sin^4\theta_W - \sin^2\theta_W + 1/4}
\textrm{Br}(Z \rightarrow l_\gamma \overline{l}_\gamma)
\, ,\nn
\eea
where we have used $|[(NN^\dagger)^2]_{\alpha\beta}
|^2=4|(NN^\dagger)_{\alpha\beta} |^2$ and $\sin^2\theta_W=0.23$ is the
Weinberg angle.
 From this,  the
bounds  in Table~\ref{branchingratios}  have been derived.
\begin{table}[!h]
\centering 
\begin{eqnarray*}
\begin{array}{|c|c|c|} 
\hline 
\displaystyle \mbox{Constraints on}  \vphantom{\frac{a}{a}}
&
\mbox{Process}
&
\mbox{Bound}
\\
\hline
\hline
\displaystyle |(NN^\dagger)_{e\mu} |
&
\displaystyle \textrm{Br} (Z \rightarrow e^\pm \mu^\mp)
&
<2.5 \cdot 10^{-3}  \rule[-8pt]{0pt}{22pt}
\\
\hline
\displaystyle |(NN^\dagger)_{e\tau} |
&
\displaystyle \textrm{Br} (Z \rightarrow e^\pm \tau^\mp)
&
<6.1 \cdot 10^{-3}   \rule[-8pt]{0pt}{22pt}
\\
\hline
\displaystyle |(NN^\dagger)_{\mu\tau} |
&
\displaystyle \textrm{Br} (Z \rightarrow \mu^\pm \tau^\mp)
&
<6.7 \cdot 10^{-3}    \rule[-8pt]{0pt}{22pt}
\\
\hline
\end{array} 
\end{eqnarray*} 
\caption{
\label{branchingratios} \it Constraints on $(NN^\dagger)_{\alpha\beta}$ from tree-level $Z$ decays into charged leptons.} 
\end{table}
%
\subsubsection{$\boldsymbol{\mu \rightarrow eee}$ and $\boldsymbol{\tau \rightarrow 3 l}$ decays}
\label{4charged}
The presence of flavour changing neutral currents in the charged
lepton sector results, in the case of the fermionic triplet Seesaw
theory under study, in tree-level $\mu\rightarrow 3e$ transitions
given by (at leading order in $\epsilon^\Sigma$):
\bea
\textrm{Br}(\mu^- \rightarrow e^+ e^- e^-)&\simeq&
\frac{\Gamma(\mu^- \rightarrow e^+ e^- e^-)}
{\Gamma(\mu^- \rightarrow e^- \nu_\mu \overline{\nu}_e)}\\
&=& |[(NN^\dagger)^2]_{e \mu} |^2 
\left( 3\sin^4\theta_W - 2\sin^2\theta_W + \frac{1}{2} \right)\,.
\nn
\eea
Analogously, $\tau$ decays in 3$e$ or 3$\mu$ are non-zero and given by:
\bea
\textrm{Br}(\tau^- \rightarrow l_\alpha^+ l_\alpha^- l_\alpha^-)&=&
\frac{\Gamma(\tau^- \rightarrow l_\alpha^- l_\alpha^+ l_\alpha^-)}
{\Gamma(\tau^- \rightarrow e^- \nu_\tau \overline{\nu}_e)}
\textrm{Br}(\tau^- \rightarrow e^- \nu_\tau \overline{\nu}_e)=\\
&=& |[(NN^\dagger)^2]_{\alpha \tau} |^2 
\left( 3\sin^4\theta_W - 2\sin^2\theta_W + \frac{1}{2} \right)
\textrm{Br}(\tau^- \rightarrow e^- \nu_\tau \overline{\nu}_e)\,,
\nn
\eea
where $\alpha = \mu , e$  
. On the
other side, $\tau$ decays in $2e(\mu)+1\mu (e)$ are given by:
\bea
\textrm{Br}(\tau^- \rightarrow l_\alpha^+ l_\alpha^- l_\beta^-)&=&
\frac{\Gamma(\tau^- \rightarrow l_\alpha^+ l_\alpha^- l_\beta^-)}
{\Gamma(\tau^- \rightarrow l_\delta^- \nu_\tau \overline{\nu}_e)}
\textrm{Br}(\tau^- \rightarrow l_\delta^- \nu_\tau \overline{\nu}_e)=\\&=&  |[(NN^\dagger)^2]_{\beta \tau}|^2 
\left(2\sin^4\theta_W-\sin^2\theta_W+\frac{1}{4}\right)
\textrm{Br}(\tau^- \rightarrow l_\delta^- \nu_\tau \overline{\nu}_e)\,,\nn\\[0.2cm]
\textrm{Br}(\tau^- \rightarrow l_\beta^+ l_\alpha^- l_\alpha^-)&=&
\frac{\Gamma(\tau^- \rightarrow l_\beta^+ l_\alpha^- l_\alpha^-)}
{\Gamma(\tau^- \rightarrow e^- \nu_\tau \overline{\nu}_e)}
\textrm{Br}(\tau^- \rightarrow e^- \nu_\tau \overline{\nu}_e)=\\
&=& \frac{1}{2}|[(NN^\dagger)^2]_{\alpha \tau}|^2 
|[(NN^\dagger)^2]_{\alpha \beta}|^2
\textrm{Br}(\tau^- \rightarrow e^- \nu_\tau \overline{\nu}_e)\,,\nn
\eea
where $\alpha ,\beta = \mu , e$ with $\alpha \neq \beta$. The bounds resulting from these processes for combinations of
$NN^\dagger$ elements are contained in Table \ref{mueee}.
\begin{table}[!ht]
\centering 
\begin{eqnarray*}
\begin{array}{|c|c|c|} 
\hline  
\displaystyle \mbox{Constraints on}  \vphantom{\frac{a}{a}}
&
\mbox{Process}
&
\mbox{Bound}
\\
\hline
\hline
\displaystyle |(NN^\dagger)_{e \mu} |
&
\displaystyle \mu^- \rightarrow e^+ e^- e^-
&
<1.1\cdot 10^{-6}   \rule[-8pt]{0pt}{22pt}
\\
\hline
\displaystyle|(NN^\dagger)_{e \tau } |
&
\displaystyle \tau^- \rightarrow e^+ e^- e^-
&
<1.2\cdot 10^{-3}   \rule[-8pt]{0pt}{22pt}
\\
\hline
\displaystyle|(NN^\dagger)_{\mu \tau} | 
&
\displaystyle \tau^- \rightarrow \mu^+ \mu^- \mu^-
&
<1.2\cdot 10^{-3}   \rule[-8pt]{0pt}{22pt}
\\
\hline
\displaystyle |(NN^\dagger)_{\tau e}|
&
\displaystyle \tau^- \rightarrow \mu^+ \mu^- e^-
&
<1.6\cdot 10^{-3}    \rule[-8pt]{0pt}{22pt}
\\
\hline
\displaystyle |(NN^\dagger)_{\tau\mu}| |(NN^\dagger)_{e \mu}|
&
\displaystyle \tau^- \rightarrow e^+ \mu^- \mu^-
&
<3.1\cdot 10^{-4}    \rule[-8pt]{0pt}{22pt}
\\
\hline
\displaystyle |(NN^\dagger)_{\tau\mu}|
&
\displaystyle \tau^- \rightarrow e^+ e^- \mu^- 
&
<1.5\cdot 10^{-3}    \rule[-8pt]{0pt}{22pt}
\\
\hline
\displaystyle |(NN^\dagger)_{\tau e}| |(NN^\dagger)_{\mu e}|
&
\displaystyle \tau^- \rightarrow  \mu^+e^-  e^-
&
<2.9 \cdot 10^{-4}    \rule[-8pt]{0pt}{22pt}
\\
\hline
\hline
\displaystyle |(NN^\dagger)_{e \mu} |   &  \mu \rightarrow e \gamma 
& 1.1 \cdot 10^{-4}  \rule[-8pt]{0pt}{22pt}\\
\hline
\displaystyle |(NN^\dagger)_{\mu \tau} |   &  \tau \rightarrow \mu \gamma 
& 1.9 \cdot 10^{-2}  \rule[-8pt]{0pt}{22pt}\\
\hline
\displaystyle |(NN^\dagger)_{e \tau} |  &  \tau \rightarrow e \gamma 
&  2.4 \cdot 10^{-2}   \rule[-8pt]{0pt}{22pt}\\
\hline
\end{array} 
\end{eqnarray*} 
\caption{
\label{mueee} \it Constraints on $(NN^\dagger)_{\alpha\beta}$ from charged leptons decays.} 
\end{table}
\subsubsection{Complete Lagrangian and $\boldsymbol{l_1} \rightarrow \boldsymbol{l_2 \gamma}$}
 
 As the phenomenological consequences of Seesaw scenarios mediated by $SU(2)$ fermionic triplets remain almost unexplored in the literature, it is worth to study in detail the complete Lagrangian for the high-energy theory
 in Eq.~(\ref{Lfermtrip}), developing it in terms of the electrically charged components\footnote{Note that the charged conjugate of 
 $\Sigma_R^\pm$  is not $\Sigma_R^\mp$ but  $\Sigma_R^{\pm c}$.
 } of $\vec\Sigma$, 
\begin{eqnarray}
\label{Lfull-ft-1}
{\cal L}&=& \overline{\Sigma_R^+} i \dv  \Sigma^+_R  +
\overline{\Sigma_{R}^-}  i\dv  \Sigma^-_R   +
\overline{\Sigma_R^0} i \dv  \Sigma^0_R  \nonumber \\
&+&g \left(W_\mu^+ \overline{\Sigma^0_R} \gamma_\mu \Sigma_R^- 
- W_\mu^+   \overline{\Sigma_R ^+} \gamma_\mu \Sigma_R^0\,+\, \text{h.c.}\right)
+g\left(W_\mu^3   \overline{\Sigma_R ^+} \gamma_\mu \Sigma_R^+ 
-W_\mu^3    \overline{\Sigma_R ^-} \gamma_\mu \Sigma_R^- \right)\nonumber\\
&-&\frac{1}{2}\left( \overline{\Sigma_R^{+}} M_\Sigma  \Sigma_R^{-c} +  \overline{\Sigma_R^{-}} M_\Sigma  \Sigma_R^{+c} + \overline{\Sigma_R ^{0}} M_\Sigma  \Sigma_R^{0c}  \,+\, \text{h.c.} \right)  \nonumber \\ 
&-&  \left( \phi^0 \overline{\Sigma_R ^0}  Y_\Sigma \nu_{L}+ \sqrt{2}\phi^0 \overline{\Sigma_R^-} Y_\Sigma l_{L}
+     \phi^+ \overline{\Sigma_R ^0}Y_\Sigma l_{L}- \sqrt{2}\phi^+ \overline{\Sigma_R ^+} Y_\Sigma \nu_{L}  \,+\, \text{h.c.}\right)\,.
\end{eqnarray}
This Lagrangian, in which the charged components of the triplets are expressed in terms of 2-component fields, is not convenient when considering mixing with the charged leptons, which as usual are expressed in 4-component notation. As the charged triplet components have 4 degrees of freedom they can all be written in terms of a 4-component unique Dirac spinor, 
\begin{equation}
 \label{Psi}
\Psi\equiv\Sigma_R^{+ c} + \Sigma_R^-\,.
\end{equation}
The neutral fermionic triplet components on the other hand can be left in 2-component notation, since they have only two degrees of freedom and mix with the neutrinos, which are also described by 2-component fields. This leads to the Lagrangian
 \begin{eqnarray}
 \label{Lfull-ft-2}
{\cal L}&=& \overline{\Psi} i \dv \Psi  + \overline{\Sigma_R^0} i \dv  \Sigma^0_R  
-  \overline{\Psi}M_\Sigma \Psi -
        \left( \overline{\Sigma^{0}_R} \frac{{M_\Sigma}}{2}  \Sigma_R^{0c} \,+  \,\text{h.c.}\right) 
\nonumber \\
&+&g \left(W_\mu^+ \overline{\Sigma_R^0} \gamma_\mu  P_R\Psi 
 +  W_\mu^+ \overline{\Sigma_R^{0c}} \gamma_\mu  P_L\Psi   \,+  \,\text{h.c.}
 \right) - g\, W_\mu^3 \overline{\Psi} \gamma_\mu  \Psi 
 \nonumber\\
\nonumber \\ 
&-&  \left( \phi^0 \overline{\Sigma_R^0} Y_\Sigma \nu_{L}+ \sqrt{2}\phi^0 \overline{\Psi} Y_\Sigma l_{L}
+     \phi^+ \overline{\Sigma_R ^0} Y_\Sigma l_{L} - \sqrt{2}\phi^+
 \overline{{\nu_{L}}^c}
Y^{T}_\Sigma \Psi   \,+  \,\text{h.c.}\right)\,.
\end{eqnarray}
The mass term of the charged sector shows 
then the usual aspect for Dirac particles (omitting flavor indices):
\begin{equation}
{\cal L} \owns -(\overline{l_R}\,\, \overline{\Psi_R} )
\,\,
\left(
\begin{array}{ cc}
   m_l  &   0 \\
      {Y_\Sigma} v &  {M_\Sigma} 
\end{array}
\right) \,\,
\left(
\begin{array}{ c}
   l_L \\
  \Psi_L 
\end{array}
\right) \,\,- 
(\overline{l_L}\,\, \overline{\Psi_L} )
\,\,
\left(
\begin{array}{ cc}
   m_l  &   Y_\Sigma^\dagger v \\
     0 &  {M_\Sigma} 
\end{array}
\right) \,\,
\left(
\begin{array}{ c}
   l_R \\
  \Psi_R \,
\end{array}
\right)\,, 
\label{chargedfullmassmatrix}
\end{equation}
The -symmetric- mass matrix for the neutral states is on the other hand given by:
\bea
{\cal L}& \owns& -(\overline{\nu_L}\,\, \overline{\Sigma^{0c}} )
\left(
\begin{array}{ cc}
  0  &   {Y_\Sigma}^\dagger v/2\sqrt{2} \\
   {Y_\Sigma}^* v/2\sqrt{2} &  {M_\Sigma}/2 
\end{array}
\right) 
\left(
\begin{array}{ c}
   \nu_L^c \\
   \Sigma^0 
\end{array}
\right) \, 
\nn\\
&&-(\overline{\nu_L^c}\,\, \overline{\Sigma^{0}} )\left(
\begin{array}{ cc}
  0  &   {Y_\Sigma}^T v/2\sqrt{2} \\
   {Y_\Sigma} v/2\sqrt{2} &  {M_\Sigma}/2 
\end{array}
\right) 
\left(
\begin{array}{ c}
   \nu_L \\
   \Sigma^{0c}
\end{array}
\right)\, .
\label{neutralfullmassmatrix}
\eea
The corresponding mixing matrices, necessary 
to calculate $\mu \rightarrow e \gamma$ and similar processes, are explicitly given in Appendix B.
 
 \subsubsection*{$\boldsymbol{\mu}\rightarrow \boldsymbol{e \gamma}$, $\boldsymbol{\tau}\rightarrow \boldsymbol{e \gamma}$ and $\boldsymbol{\tau}\rightarrow \boldsymbol{\mu \gamma}$}
 
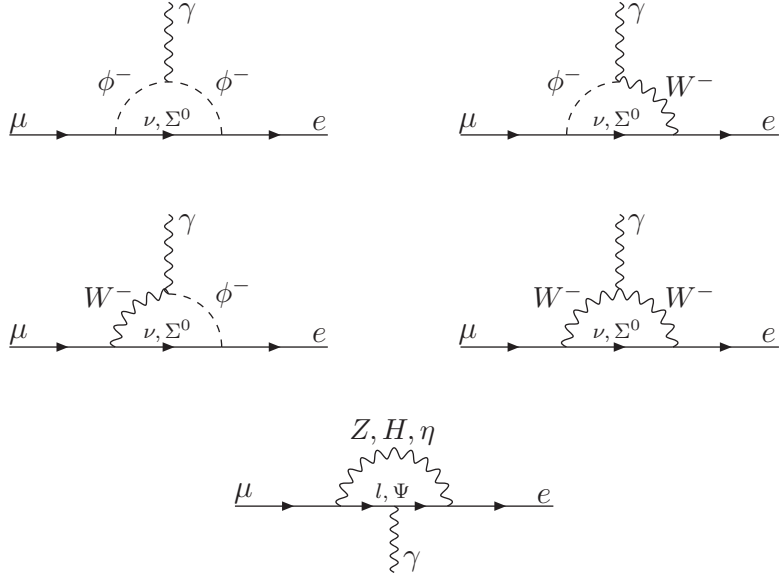
\begin{figure}[t]
\centering
\vspace{15mm}
\begin{picture}(330,150)(0,-100)
\ArrowLine(10,80)(50,80)
\ArrowLine(50,80)(90,80)
\ArrowLine(90,80)(130,80)
\DashCArc(70,80)(20,0,180){3}
\Photon(70,100)(70,130){1.5}{5}
\Text(10,81)[bl]{$\mu$}
\Text(130, 82)[br]{$e$}
\Text(74,130)[tl]{$\gamma$}
\Text(57,95)[br]{\small{$\phi^-$}}
\Text(88,95)[bl]{\small{$\phi^-$}}
\Text(70,82)[b]{\scriptsize{$\nu,\Sigma^0$}}
\ArrowLine(180,80)(220,80)
\ArrowLine(220,80)(260,80)
\ArrowLine(260,80)(300,80)
\DashCArc(240,80)(20,90,180){3}
\PhotonArc(240,80)(20,0,90){2}{5.5}
\Photon(240,100)(240,130){1.5}{5}
\Text(180,81)[bl]{$\mu$}
\Text(300, 82)[br]{$e$}
\Text(244,130)[tl]{$\gamma$}
\Text(227,95)[br]{\small{$\phi^-$}}
\Text(258,95)[bl]{\small{$W^-$}}
\Text(240,82)[b]{\scriptsize{$\nu,\Sigma^0$}}
\ArrowLine(10,0)(50,0)
\ArrowLine(50,0)(90,0)
\ArrowLine(90,0)(130,0)
\DashCArc(70,0)(20,0,90){3}
\PhotonArc(70,0)(20,90,180){2}{5.5}
\Photon(70,20)(70,50){1.5}{5}
\Text(10,1)[bl]{$\mu$}
\Text(130, 2)[br]{$e$}
\Text(74,50)[tl]{$\gamma$}
\Text(57,15)[br]{\small{$W^-$}}
\Text(88,15)[bl]{\small{$\phi^-$}}
\Text(70,2)[b]{\scriptsize{$\nu,\Sigma^0$}}
\ArrowLine(180,0)(220,0)
\ArrowLine(220,0)(260,0)
\ArrowLine(260,0)(300,0)
\PhotonArc(240,0)(20,0,180){2}{10.5}
\Photon(240,22)(240,50){1.5}{5}
\Text(180,1)[bl]{$\mu$}
\Text(300,2)[br]{$e$}
\Text(244,50)[tl]{$\gamma$}
\Text(227,15)[br]{\small{$W^-$}}
\Text(258,15)[bl]{\small{$W^-$}}
\Text(240,2)[b]{\scriptsize{$\nu,\Sigma^0$}}
\ArrowLine(95,-60)(135,-60)
\ArrowLine(135,-60)(155,-60)
\ArrowLine(155,-60)(175,-60)
\ArrowLine(175,-60)(215,-60)
\PhotonArc(155,-60)(20,0,180){2}{10.5}
\Photon(155,-60)(155,-85){1.5}{5}
\Text(95,-59)[bl]{$\mu$}
\Text(215, -59)[br]{$e$}
\Text(159,-85)[bl]{$\gamma$}
\Text(155,-37)[b]{\small{$Z,H,\eta$}}
\Text(155,-59)[b]{\scriptsize{$l,\Psi$}}
\end{picture}
\caption{Diagrams contributing to $\mu\rightarrow e \gamma$. $\phi^-$
is the Goldstone boson associated with the $W^-$ boson, $\eta$ is
the Goldstone boson associated with the Z boson and $H$ stands for the physical Higgs boson.}
\label{muegammatypeIII}
\end{figure}
$l_1 \rightarrow l_2 \gamma$ transitions result  from $Z$- and $W$-mediated one-loop processes, depicted in Fig.~\ref{muegammatypeIII}. The amplitude of the matrix element, computed within the complete theory,  Eqs.~(\ref{Lfermtrip}) and (\ref{Lfull-ft-2}),  is given by (the details of the computation will be given in a separate publication~\cite{Abada:2008ea}):\begin{eqnarray}
\mathcal{A}_{l_1\rightarrow l_2\gamma}&=&-i\frac{G_F^{SM}}{\sqrt{2}}\frac{e}{16\pi^2}m_1\overline{u_2}\left(p-q\right)P_R i\sigma_{\lambda\nu}q^{\nu}\epsilon^{\lambda}u_1\left(p\right)\times\nonumber\\
&&\left\{C\,\epsilon^{\Sigma}_{21}+\sum_i x_{\nu_i}\left(U_{0_{\nu\nu}}\right)_{2i}\left(\left(U_{0_{\nu\nu}}\right)^{\dagger}\right)_{i1}+{\cal O}\left(\frac{1}{M_{\Sigma}^4}\right)\right\}
\end{eqnarray}
In this equation, $C=2.23$, $x_{\nu_i}=\frac{m^2_{\nu_i}}{M^2_W}$ and  $\epsilon^\Sigma_{21}$ corresponds to the $d=6$ operator coefficients $\epsilon^\Sigma_{e\mu}$, $\epsilon^\Sigma_{e\tau }$ and $\epsilon^\Sigma_{\mu\tau}$  in Eq.~(\ref{epsilonfermtrip}), when considering ${ \mu\rightarrow e \gamma}$, ${ \tau\rightarrow e \gamma}$ and ${\bf \tau\rightarrow \mu \gamma}$ transitions, respectively.
$U_{0_{\nu\nu}}$ is the unitary matrix which diagonalizes the 
 neutral lepton mass matrix for the fields $(\nu_L, \Sigma^{0c})$, see Appendix B for details. 
Using these results, and Eq.~(\ref{G_F_fermtrip}) the branching ratio  for the  $l_1 \rightarrow l_2\gamma$ transition   is given by (at order $1/M_{\Sigma}^2$):
\begin{equation}
BR\left(l_1\rightarrow l_2\gamma\right)=\frac{3}{32}\frac{\alpha}{\pi}\frac{\left|C\,\epsilon^{\Sigma}_{21}+\sum_i x_{\nu_i}\left(U_{0_{\nu\nu}}\right)_{2 i}\left(\left(U_{0_{\nu\nu}}\right)^{\dagger}\right)_{i1}\right|^2}{(NN^\dagger)_{11} (NN^\dagger)_{22}}
\end{equation}
 The experimental bounds on these processes result in
  constraints given in Table 8. These are comparable to those stemming from tree-level purely leptonic decays.
  
%
\subsubsection{Combination of all constraints}
\noindent
From all constraints obtained above we have performed a global fit, 
and the following bounds on the
$NN^\dagger$ elements have been derived, at the 90\% CL:
\begin{eqnarray}
|N N^\dagger | \approx
\begin{pmatrix}
1.001 \pm 0.002   & <1.1\cdot 10^{-6}  & <1.2\cdot 10^{-3} \\
<1.1\cdot 10^{-6}  & 1.002 \pm 0.002  &  <1.2\cdot 10^{-3} \\
<1.2\cdot 10^{-3}  &  <1.2\cdot 10^{-3}  & 1.002 \pm 0.002 
\end{pmatrix} \, .
\label{nndag-bis}
\end{eqnarray}
Using now the relation obtained in  Eq.~(\ref{Nc6-ft}) between the
elements of the coefficient matrix $c^{d=6}$ and those of
$NN^\dagger$, it follows that 
\begin{eqnarray}
\label{bounds-III}
\frac{v^2}{2}\,|c^{d=6}|_{\alpha\beta}\,=\,\frac{v^2}{2}\,|Y_\Sigma^\dagger\frac{1}{M_\Sigma^\dagger}\frac{1}{ M_\Sigma}Y_{\Sigma}|_{\alpha\beta}\ \lesssim
\begin{pmatrix}
3 \cdot 10^{-3}   & <1.1\cdot 10^{-6}  & <1.2\cdot 10^{-3} \\
<1.1\cdot 10^{-6}   & 4\cdot 10^{-3}  &  <1.2\cdot 10^{-3} \\
<1.2\cdot 10^{-3}   &  <1.2\cdot 10^{-3}  & 4\cdot 10^{-3}
\end{pmatrix}  .
\label{limitsIII}
\end{eqnarray}
Notice that these bounds are stronger than those obtained in the
case of the fermionic singlet Seesaw theory, Eq.~(\ref{limits}). This is due to the fact that now
flavour changing processes with charged fermions are allowed already at
tree level.

\subsubsection{Signals at colliders from fermionic triplets}
As for direct production and detection, alike to the case of the
generic type-II Seesaw model, the non-zero electroweak charge of the
triplet results in gauge production from photon and Z couplings. Only
particles with electric charge $\pm 1$ exist in this case, though, and
the experimental signals are less clean. Anyway, if light enough,
triplet fermions can be produced in forthcoming colliders through
Drell-Yan production. In Ref.~\cite{Ma:2002pf,bajc}, the following channels have
been analyzed:
\begin{itemize}
\item $\Sigma$ decays into gauge bosons plus light leptons:
$\Sigma^-\rightarrow Z l^-$, $\Sigma^-\rightarrow W^- \nu$,
$\Sigma^0\rightarrow Z \nu$, $\Sigma^0\rightarrow W^\pm l^\mp$;
\item $\Sigma$ decays into Higgs plus light leptons:
$\Sigma^-\rightarrow \phi^0 l^-$, $\Sigma^0\rightarrow \phi^0 \nu$.
\end{itemize} 
 
\newpage 
\section{Conclusions}
\label{conclusions}
While the unique dimension five effective operator is common to all 
Seesaw models of Majorana neutrinos,  dimension six operators discriminate among them. We have  determined the latter for the three families of Seesaw models: fermionic singlet (typeI), scalar triplet (type II) and fermionic triplet (type III). They should be the low-energy tell-tale of the Seesaw mechanism, for any generic  beyond the SM theory whose typical scale is larger than the electroweak scale and which  accounts for Majorana neutrino masses. These results have been gathered in Table 1. 
   
For fermionic Seesaw theories, the effective operators obtained  result in  non-unitary leptonic mixing matrices affecting the couplings of leptons to gauge bosons, in very precise patterns. Denoting by $N$ the non-unitary matrix which replaces the usual $U_{PMNS}$ matrix in the charged current, the neutrino-$Z$ and charged lepton-$Z$ currents have now a flavour structure given by
       \begin{eqnarray}
    &J_\mu^{3-\nu}&\propto
     N^\dagger N\,,\,\,\,\,\,\,\,\,\,\,\,\,\,
     J_\mu^{3-l}\propto
     1\,,\,\,\,\,\,\,\,\,\,\,\,\,\,\,\,\,\,\,\,\,\,\, \text{for\, singlet-fermion\,Seesaw}\,, \nonumber\\ 
   &J_\mu^{3-\nu}&\propto
     (N^\dagger N)^{-1}\,,\,\,\,\,
     J_\mu^{3-l}\propto
     (NN^\dagger)^2\,,\,\,\,\, \text{for\, triplet-fermion\,Seesaw}\,. \nonumber 
       \end{eqnarray}
   For scalar-triplet Seesaw theories the mixing matrices remain unitary, while the dimension six operators indicate instead correlations between exotic four-fermion couplings and gauge and Higgs potential parameters, as well as with  Higgs transitions.
   
   For all families of Seesaw theories, it turns out that the  coefficient matrix of the dimension six leptonic  operators is of the generic form $|c^{d=6}|= Y^\dagger \frac{1}{M^2} Y$, where $Y$ denote the new Yukawa couplings and $M$ the high scale of the new theory. Irrespective of the value of $M$, we have set bounds on the $Y/M$ ratios   for the three theories, resulting in an overall constraint $|Y|\lesssim 10^{-1} \frac {M}{TeV}$, with more stringent constraints for specific channels, specially for type II and III theories due the richness of their phenomenology. The specific results have been collected in Eq.~(\ref{bounds-I}), Tables 
 4 and 5, and Eq.~(\ref{bounds-III}), for the fermionic singlet, scalar triplet and fermionic triplet Seesaw theories, respectively.
To achieve them, we took into account the experimental data on many tree-level processes as well as on radiative one-loop processes ($\mu\rightarrow e \gamma$, $\tau \rightarrow \mu \gamma$ and $\tau \rightarrow e \gamma$),
  all of which we computed  in all 3 theories.
  
  Independently of the above, we have also discussed the possible values of $M$ from a theoretical, albeit model independent, point of view. 
There is no clue at present on whether there is a relationship between the source of $B-L$ violation in nature and the origin of the flavour structure of the SM, which displays Yukawa couplings for charged fermions ranging from $Y\sim 1$ for the top quark to $\sim 10^{-6}$ for the electron. The values of neutrino Yukawa couplings could also be in that same range within  Seesaw theories, if the high energy scale lies in the range between the typical  Grand Unification scale down to the TeV scale. 
Indeed, the electroweak hierarchy problem prefers new physics scales closer to the electroweak scale than to the hypothetical Grand Unified scale, if the new physics involves the Higgs field.
To illustrate this point in the present context, we have explicitly computed the one-loop contributions to the Higgs mass in the three families of Seesaw theories, showing its quadratic sensitivity to the new scales.
We have then also addressed in this work the question of  whether it is possible to simultaneously allow a ``low" scale $M\sim TeV$ and large, order one, Yukawa couplings, without fine-tuning neither any Yukawa coupling nor combinations of them. 
The answer is positive and guided by symmetry considerations. Indeed, while neutrino masses correspond to the dimension five operator which violates $B-L$, all dimension six operators preserve it.  From the point of view of symmetries, it is then a sensible option to expect large effects of the new physics associated to the latter, while the dimension five operator is further suppressed. 
A natural ansatz proposed is that, if in the new theory 
the Majorana character  is associated to some tiny parameter $\mu$ which heralds the breaking of $B-L$, the dimension five operator coefficient is necessarily proportional to it, $c^{d=5}\sim \frac{\mu}{M^2}$ and thus suppressed. This mechanism and pattern is stable under radiative corrections, as they have to be proportional to $\mu$, which is the parameter responsible for the small breaking of a global symmetry. We call this pattern {\it direct lepton violation} since the neutrino masses are proportional to the (small) lepton number violating quantity $\mu$, rather than inversely proportional to the large
lepton number violating heavy field mass.
 It turns out that such an ansatz and pattern is already incorporated in the minimal scalar-triplet Seesaw theory (type II). We have also argued that fermionic Seesaw theories at the TeV scale (as for instance the so-called inverse Seesaw mechanism)  include it  as well and we have explored the corresponding possible textures and realizations. 
Would this ansatz happen in nature, new beautiful signals may be expected near the present experimental bounds and in accelerators sensitive to the TeV scale, such as the LHC or ILC . 
The loose bounds we have obtained for the dimension six operator coefficients, above mentioned, show that it is indeed possible to have such strong signals, with $M\sim$ TeV, Yukawa couplings of ${\cal O}(10^{-2}$-$1)$ and no unnatural fine-tunings.
\section*{Acknowledgments}
We specially thank Stefan Antusch, who participated in the early stages of this work. We also
acknowledge illuminating discussions with Jos\'e Ram\'on Espinosa, Enrique Fern\'andez-Mart\'{\i}nez, Jacobo L\'opez-Pav\'on  and Stefano Rigolin. 
Furthermore, the authors received partial
support from CICYT through the project FPA2006-05423, as well as from
the Comunidad Aut\'onoma de Madrid through Proyecto HEPHACOS;
P-ESP-00346.  T.H. thanks the FNRS-FRS for support. A.A and F.B aknowledge the support of the Agence Nationale de la Recherche ANR through the project JC05-43009-NEUPAC.
\newpage
\section{Appendix A:\,\,Non-unitarity}
\label{AppN}
We give here in detail the transformations leading to the currents in Eqs.~(\ref{JCC_d6}), (\ref{JNC_d6}) and Eqs.~(\ref{JCC_fermtrip}), (\ref{JNCnu_fermtrip}), (\ref{JNCe_fermtrip}),  corresponding to the singlet and triplet fermionic Seesaws, respectively.
\subsection*{Singlet fermion Seesaw}
Consider the Lagrangians in Eqs.~(\ref{LagShift}) and (\ref{JI}).
We can  rotate to the basis in which 
the mass matrices are diagonal. 
In this basis, the neutrino light eigenstates are redefined as
\begin{eqnarray}
\nu_i&=& V_{i \alpha }^{\rm{eff}} \,{\nu_\alpha}_L
+ V^{\rm{eff}\,*}_{i \alpha }\,{\nu_\alpha}^c_L~,\\
\end{eqnarray}
where
$V^{\rm eff}$  are not  unitary matrices because of the field rescaling involved.  $V^{\rm eff}$ can be expressed in terms of the matrix which diagonalizes the neutrino mass matrix\footnote{ Notice that  $U^\nu$ does not depend on $c^{d=6}$ at ${\cal O}(1/M^2)$.} $U^\nu$, 
\begin{eqnarray}
 V^{\rm{eff}}= (1-\frac{1}{2}\epsilon^{N})\,U^\nu\,.\\
  \end{eqnarray}
In terms of the light mass eigenstates,  the
leptonic Lagrangian now becomes
\bea
{\cal L}_{\rm leptons}^{d\le 6}=
\frac{1}{2}\overline {\nu_i} \left( i\dv- m^{diag}_{\nu\,i} \right) {\nu_i}+\frac{1}{2}\overline {l_i} \left( i\dv- m^{diag}_{l\,i} \right) {l_i}+ \mathcal{L}_{CC} + \mathcal{L}_{NC}\,,
\eea
where, in this mass basis, the charged and neutral currents 
read
\begin{eqnarray}
\mathcal{L}_{CC} &=&\frac{g}{\sqrt{2}}\overline{l_{L}}{W\!\!\!\!\!/}\;^{-}\left[\Omega\, \left(1-\frac{1}{2}\epsilon^{N}\right)\,U^\nu\,\right]{\nu_{L}}+ \text{h.c.}\,,\\
\mathcal{L}_{NC}&=&\frac{g}{cos\theta_{W}}\left\{\frac{1}{2}\left[\overline{\nu_{L}} \gamma_{\mu}\left[{U^\nu}^\dagger\left(1-\epsilon^{N} \right)\,U^\nu\,\right]{\nu_{L}}-\overline{l_{L}}\gamma_{\mu}{l_{L}}\right]-sin^{2}\theta_{W}J^{em}_{\mu}\right\}Z^{\mu}\nonumber \\
&+&eJ^{em}_{\mu}A^{\mu}\,,
\end{eqnarray}
where $\Omega\equiv{\rm diag}(e^{i\omega_1},e^{i\omega_2},e^{i\omega_3})$ reabsorbs three unphysical phases in the definition of the charged lepton fields. The above expressions look quite complicated, but the measurable effects can be expressed in a compact way,  
 denoting by N the non-unitary matrix appearing in the charged current coupling,
\bea
\label{NNappendix}
N\equiv \Omega\, \left(1-\frac{1}{2}\epsilon^{N}\right)\,U^\nu\,.
\eea
In terms of the matrix $N$, the charged and neutral currents  read
\bea
\label{JCC_d6-bis}
J_\mu^{-\,CC}  
&\equiv& \overline {e_L}_\alpha \, \gamma_\mu \,
N_{\alpha i} \, \nu_i,\\
\label{JNC_d6-bis}
J_\mu^{NC}
&\equiv& {1 \over 2} \overline \nu_i \,\gamma_\mu
(N^\dagger \,N)_{i j}
\,\nu_j ,
\eea
with ${N_{i \alpha}}^\dagger\,N_{\alpha j} \ne
\delta_{ij}$ appearing in the neutral current since $N$ is not
unitary.
\subsection*{Triplet fermion Seesaw}
Consider now the Lagrangians in Eqs.~(\ref{LagShift2})-(\ref{LagCCNCfermtrip}).
We can  rotate to the basis in which both the leptonic kinetic energies and 
their mass matrices are diagonalized. 
This requires to redefine the leptonic light fields as
\begin{eqnarray}
\nu_i&=& V_{i \alpha }^{\rm{eff}} \,{\nu_\alpha}_L
+ V^{\rm{eff}\,*}_{i \alpha }\,{\nu_\alpha}^c_L~,\\
{l_L}_i &=& K_{i \alpha }^{\rm{eff}} \, {l_L}_\alpha ,\\
{l_R}_i &=&( {U^l_R}) _{i \alpha }\,{l_R}_\alpha, 
\end{eqnarray}
where $U^l_R$ is unitary while $V^{\rm eff}$ and $K^{\rm eff}$ are not  unitary matrices because of the field rescaling involved,
\begin{eqnarray}
 V^{\rm{eff}}= (1-\frac{1}{2}\epsilon^{\Sigma})\,U^\nu\\
 K^{\rm{eff}}=(1-\epsilon^{\Sigma})\,U^l_L\,,
  \end{eqnarray}
 with the matrices $U_\nu$ and $U^l$ diagonalizing the
 neutrino  and charged lepton mass terms, respectively\footnote{ While $U^\nu$ does not depend on $c^{d=6}$ at ${\cal O}(1/M^2)$, $U^l_L$ and $U^l_R$ do.}, 
  with
  \begin{eqnarray}
  m_\nu^\text{diag}\equiv {U^\nu}^T \, m_\nu\,U^\nu\,,\hspace{1.5cm}
  m_l^\text{diag}\equiv {U^l_R}^\dagger \, m_l\, (1 - \epsilon)\,U^l_L\,.  \end{eqnarray}
In terms of the light mass eigenstates,  the
leptonic Lagrangian becomes
\bea
{\cal L}_{\rm leptons}^{d\le 6}=
\frac{1}{2}\overline {\nu_i} \left( i\dv- m^{diag}_{\nu\,i} \right) {\nu_i}+\frac{1}{2}\overline {l_i} \left( i\dv- m^{diag}_{l\,i} \right) {l_i}+ \mathcal{L}_{CC} + \mathcal{L}_{NC}\,,
\eea
 in which the charged and neutral currents now  are given by
\begin{eqnarray}
\mathcal{L}_{CC} &=&\frac{g}{\sqrt{2}}\overline{l_{L}}{W\!\!\!\!\!/}\;^{-}\left[\Omega\,{U^l_L}^\dagger \left(1+\frac{1}{2}\epsilon^{\Sigma}\right)\,U^\nu\,\right]{\nu_{L}}+ \text{h.c.}\,,\\
\mathcal{L}_{NC}&=&\frac{g}{cos\theta_{W}}\left\{\frac{1}{2}\left[\overline{\nu_{L}} \gamma_{\mu}\left[{U^\nu}^\dagger\left(1-\epsilon^{\Sigma} \right)\,U^\nu\,\right]{\nu_{L}}-\overline{l_{L}}\gamma_{\mu}\left[\Omega\,{U^l_L}^\dagger \left(1+2\epsilon^{\Sigma}\right)U^l_L\Omega^\dagger\right]{l_{L}}\right]\right.\nn\\
&-& sin^{2}\theta_{W}J^{em}_{\mu}\Big\}Z^{\mu}
+eJ^{em}_{\mu}A^{\mu}\,,
\end{eqnarray}
with, once again, $\Omega\equiv{\rm diag}(e^{i\omega_1},e^{i\omega_2},e^{i\omega_3})$ reabsorbing three unphysical phases in the definition of the charged lepton fields. 
Because of the flavour-dependent field rescaling involved, a non-unitary mixing matrix $N$ has appeared in the charged-current couplings, replacing the usual unitary $U_{PMNS}$ matrix, while non-unitary flavour mixing appears as well in the couplings of leptons to the $Z$ boson.
The above expressions look quite cumbersome, but the measurable effects can be cast in a compact way. Indeed,  we denote by N the non-unitary matrix appearing in the charged current coupling,
\bea
\label{NSigma-bis}
N\equiv \Omega\,{U^l_L}^\dagger \left(1+\frac{1}{2}\epsilon^{\Sigma}\right)\,U^\nu\,.
\eea
Working at order ${\cal O} (1/M^2)$, i.e. at first order in  the $\epsilon^\Sigma$ parameters,
 the charged and neutral currents can then be neatly expressed in the mass basis as
 \bea
\label{JCC_fermtrip-bis}
J_\mu^{-\,CC}  
&\equiv& \overline {l_L} \, \gamma_\mu \,
N \, \nu,\\
\label{JNCnu_fermtrip-bis}
J_\mu^{3} (\text{neutrinos})
&\equiv& {1 \over 2} \overline \nu \,\gamma_\mu
(N^\dagger \, N)^{-1}
\,\nu\,,\\
\label{JNCe_fermtrip-bis}
J_\mu^{3} (\text{leptons})
&\equiv& {1 \over 2} \overline l \,\gamma_\mu
(NN^\dagger)^2
\,l .
\eea
\newpage
\section{Appendix B:\,\,Lepton mixing in the full type-III Seesaw model}
As the type-III model has not been properly presented in an extensive
way for what concerns notations, mass matrices, mixing matrices, etc.,
it is useful to discuss it also in the context of the full theory
where the triplets of fermions are not integrated out. This will also
allow to establish the precise tree-level connection between the
effective and full theories.  This model is defined by
Eq.~(\ref{Lfermtrip}) in a vector notation.  It can be equivalently
rewritten in terms of the usual and compact two-by-two notation for
triplets (with implicit flavour summation):
\begin{equation}
\label{Lfermtriptwobytwo}
{\cal L}=Tr [ \overline{\Sigma} i \slash \hspace{-2.5mm} D  \Sigma ] 
-\frac{1}{2} Tr [    \overline{\Sigma}  M_\Sigma \Sigma^c 
                 +   \overline{\Sigma^c}M^*_\Sigma   \Sigma     ] 
- \tilde{\phi}^\dagger \overline{\Sigma} \sqrt{2}Y_\Sigma L 
-  \overline{L}\sqrt{2} {Y_\Sigma}^\dagger  \Sigma \tilde{\phi}
\end{equation} 
with, for each  fermionic triplet, 
\begin{eqnarray}
\Sigma&=&
\left(
\begin{array}{ cc}
   \Sigma^0/\sqrt{2}  &   \Sigma^+ \\
     \Sigma^- &  -\Sigma^0/\sqrt{2} 
\end{array}
\right), \quad 
\Sigma^c=
\left(
\begin{array}{ cc}
   \Sigma^{0c}/\sqrt{2}  &   \Sigma^{-c} \\
     \Sigma^{+c} &  -\Sigma^{0c}/\sqrt{2} 
\end{array}
\right), \nonumber\\
D_\mu &=& \dv_\mu-i\sqrt{2} g \left(
\begin{array}{ cc}
   W^3_\mu/\sqrt{2}  &   W_\mu^+ \\
     W_\mu^- &  -W^3_\mu/\sqrt{2} 
\end{array}
\right)\,.
\end{eqnarray}
Either way, Eq.~(\ref{Lfermtrip}) or Eq.~(\ref{Lfermtriptwobytwo}),
lead to the same Lagrangian expressed in terms of charge components,
as given in Eq.~(\ref{Lfull-ft-1}) or, in terms of the more convenient
spinor field $\Psi$ ( Eq.~(\ref{Psi}) ), in
Eq.~(\ref{Lfull-ft-2}). This leads to the mass matrices for both
neutral and charged leptons in
Eqs.~(\ref{chargedfullmassmatrix})-(\ref{neutralfullmassmatrix}).  As
it happens with any Dirac mass matrix, the charged lepton mass matrix
can be diagonalized by a bi-unitary matrix transformation (six-by-six
if there are three triplets of fermions)\,,
\begin{equation}
\left(
\begin{array}{ c}
   l_{L,R} \\
  \Psi_{L,R} 
\end{array}
\right) = U_{L,R}
\left(
\begin{array}{ c}
   l'_{L,R} \\
   \Psi'_{L,R}
\end{array}
\right)\,,
\end{equation}
while the symmetric neutral lepton mass matrix can be diagonalized by
a single unitary matrix:
\begin{equation}
\left(
\begin{array}{ c}
   \nu_{L} \\
   \Sigma^{0c}
\end{array}
\right) = U_0
\left(
\begin{array}{ c}
   \nu'_{L} \\
   \Sigma'^{0c}
\end{array}
\right).
\end{equation}
Writing the mixing matrices in terms of three-by-three blocks 
\begin{equation}
U_{L}\equiv
\left(
\begin{array}{ cc}
   U_{Lll} &   U_{L l\Psi} \\
     U_{L\Psi l} & U_{L \Psi\Psi}  
\end{array}
\right) \,,\,
U_{R}\equiv
\left(
\begin{array}{ cc}
   U_{Rll} &   U_{R l\Psi} \\
     U_{R\Psi l} & U_{R \Psi\Psi}  
\end{array}
\right) \,,\,
U_{0}\equiv 
\left(
\begin{array}{ cc}
   U_{0 \nu \nu } &   U_{0 \nu \Sigma} \\
     U_{0 \Sigma \nu} & U_{0 \Sigma \Sigma}  
\end{array}
\right)\,. 
\end{equation}
at order ${\cal O}([(Y_{\Sigma}v, m_l)/M_\Sigma]^2)$ we obtain:
\be
\begin{array}{llll}
U_{Lll}=1-\epsilon^\Sigma & U_{L l\Psi}=Y_{\Sigma}^\dagger M^{-1}_\Sigma v &
U_{L \Psi l}=- M^{-1}_\Sigma Y_{\Sigma} v & U_{L\Psi\Psi} = 1-\epsilon' \\
U_{Rll}=1 & U_{R l\Psi}=m_l Y_{\Sigma}^{\dagger}M_{\Sigma}^{-2}  v &
U_{R \Psi l}=- M_{\Sigma}^{-2}Y_{\Sigma} m_l v & U_{R\Psi\Psi}=1 \\
U_{0\nu\nu}=(1-\frac{\epsilon^\Sigma}{2}) U_{PMNS} & 
U_{0 \nu \Sigma}= Y_{\Sigma}^\dagger M^{-1}_\Sigma \frac{v}{\sqrt{2}} &
U_{0\Sigma \nu}=-M^{-1}_\Sigma Y_{\Sigma} \frac{v}{\sqrt{2}} U_{0\nu\nu} &
U_{0 \Sigma \Sigma}=(1-\frac{\epsilon'}{2})
\label{mixfull}
\end{array}
\ee
where $\epsilon^\Sigma=\frac{v^2}{2}Y_\Sigma^\dagger M^{-2}_\Sigma
Y_\Sigma$, $\epsilon'=\frac{v^2}{2}M^{-1}_\Sigma Y_\Sigma
Y_\Sigma^\dagger M^{-1}_\Sigma$ and $U_{PMNS}$ is the lowest order
neutrino mixing matrix which is unitary.  The six-by-six mixing
matrices $U_{L,R,0}$ are unitary but the various three-by-three ones
are not. This leads to non-unitary effects in the gauge interactions
of leptons. Re-expressing the gauge interactions in the mass
eigenstate basis we get
Eqs. (\ref{JCC_fermtrip})-(\ref{JNCe_fermtrip}) with
\begin{eqnarray}
(N N^\dagger)^2&=&1+U^\dagger_{L\Psi l} U_{L\Psi l}=1+2 \epsilon^\Sigma
\label{eql}\,,\\
(N^\dagger N)^{-1}&=&1-U^\dagger_{0 \Sigma \nu} U_{0 \Sigma \nu}=1-U^\dagger_{PMNS} \,\epsilon^\Sigma \, U_{PMNS}\,,
\label{eqnu}\\
N&=&(U^\dagger_{Lll} U_{0\nu\nu} + \sqrt{2} U^\dagger_{L\Psi l} U_{0 \Sigma \nu})=\left(1+\frac{\epsilon^\Sigma}{2}\right) U_{PMNS}\,.
\label{eqW}
\end{eqnarray}
In obtaining these results recall that, in the full high-energy
theory, all the analysis has been performed in the flavour basis in
which the initial charged lepton mass matrix is diagonal and the light
charged lepton fields have reabsorbed three arbitrary phases.  In the
last equalities of Eqs.~(\ref{eql}) and (\ref{eqnu}) we have used
Eq.~(\ref{mixfull}), while the last equality of Eq.~(\ref{eqW}) can be
obtained from combining Eqs. (\ref{eql}) and (\ref{eqnu}).  The
results we get in terms of the Yukawa couplings are fully in agreement
with the ones obtained in the effective theory,
Eqs.~(\ref{JCC_fermtrip})-(\ref{NSigmabasis}) and
Eq.~(\ref{Nc6-ft}). Note that in Eqs.~(\ref{mixfull})-(\ref{eqnu})
(although not in~Eq.~(\ref{eqW})), $U_{PMNS}$ can be replaced by $N$
since the difference is of higher order in $Yv/M_\Sigma$.

\newpage
\section{Appendix C:\,\, Low scale models of light neutrino mas\-ses with large Yukawa couplings}
\label{App3}
In the following, we consider models based on type-I 
Seesaw mechanism which lead to large dimension 6 operators.
The examples considered can be straightforwardly applied to the type-III Seesaw too, as the textures are exactly the same.
Such a situation arises for particular patterns of the singlet neutrino 
mass  matrix
and/or of the Yukawa matrix. As already explained in Sect. 3 we are interested in a class of models 
which, to lead to sufficiently suppressed neutrino masses and large  
$d=6$ operators, do not require any precise cancellations between the various (a 
priori independent) entries of these 2 matrices.
The cases we 
consider just require that some of the entries of these mass matrices are 
much smaller than other ones.
For simplicity, let us first consider - as in Sect. 3 - only one  left-handed neutrino and two 
singlet fermions. In full generality, in this case there are 3 mass 
matrix textures which 
automatically lead to a vanishing light neutrino mass [in 
the basis $(\nu_L,N_1,N_2)$]:
\begin{equation}
\left( \begin{array}{ccc}0 & m_{D_1} & 0 \\ 
m_{D_1} & M_{N_2} & M_{N_1} \\ 0 & M_{N_1} & 0
\end{array} \right) ,\quad \quad
\left( \begin{array}{ccc}0 & 0 & m_{D_1} \\ 
0 & 0 & M_{N_1} \\ m_{D_1} & M_{N_1} & M_{N_2}\end{array} \right) ,\quad \quad
\left( \begin{array}{ccc}0 & m_{D_1} & m_{D_2} \\ 
 m_{D_1} & 0 & 0 \\ m_{D_2} & 0 & 0 \end{array} \right)
\end{equation}
In the following we will consider only the first mass matrix since 
the second one is equivalent
to the first one under $N_1 \leftrightarrow N_2$, and since the third one 
which is of the Dirac type doesn't lead to any interesting case for our purposes.
Assuming $M_{N_1}> m_{D_1}$ the eigenstate which is predominantly a $\nu_L$ 
is massless. For $M_{N_2}=0$, which corresponds to the well-known 
inverse Seesaw model \cite{GonzalezGarcia:1988rw} considered in Sect.~3, this can be understood easily from 
the fact that assigning $L=1,-1,1$ to $\nu_L, N_1, N_2$ respectively, lepton 
number is conserved. For $M_{N_2}\neq 0$, which can also be justified from 
a symmetry in 
specific extended models \cite{Dudas}, this remains true 
because the determinant of the mass matrix still vanishes in this case.
This case has the interesting feature to have a large source  
of lepton number violation (i.e.~$M_{N_2}$)
with a vanishing neutrino mass~\footnote{Note that since the 22 element breaks lepton number, it could induce  
neutrino mass
in presence of extra interactions coupling to the $N_i$'s. This  
contribution would
be suppressed by loop factors, couplings of the extra interactions,  
as well as the masses of the new states involved, but wouldn't be  
necessarily negligible with respect to the contribution of Eq.~(\ref{mnu1}) below. We  
thank S. Antusch, M. Frigerio and J. Kersten for discussions on this  
point.}.
In order to induce a naturally small neutrino mass, even if the Yukawa coupling
in $m_{D_1}$ is large, and without fine-tuning, one must introduce
a small mass parameter $\mu$ in the mass matrix.
This can be done in 2 ways (plus combination of them), 
either from introducing an extra small Majorana mass, or from introducing 
an extra small Dirac mass term:
\begin{equation}
\left( \begin{array}{ccc}0 & m_{D_1} & 0 \\ 
m_{D_1} & M_{N_2} & M_{N_1} \\ 0 & M_{N_1} & \mu\end{array} \right),\quad \quad
\left( \begin{array}{ccc}0 & m_{D_1} & \mu \\ 
m_{D_1} & M_{N_2} & M_{N_1} \\ \mu & M_{N_1} & 0\end{array} \right)\ .
\label{2matrices}
\end{equation}
Expanding in powers of $\mu$, in the first case in Eq.~(\ref{2matrices}) we obtain:
\begin{equation}
m_\nu=\frac{m_{D_1}^2}{M_{N_1}}\frac{\mu}{M_{N_1}}\frac{M_{N_1}^2}{M_{N_1}^2
+m_{D_1}^2} +{\cal{O}}(m_{D_1}^2 \mu^2 M_{N_2}/M_{N_4}^4, \mu^3)\,,
\label{mnu1}
\end{equation}
while the second case leads to:
\begin{equation}
m_\nu=-2\frac{m_{D_1} \mu}{M_{N_1}}\frac{M_{N_1}^2}{M_{N_1}^2+m_{D_1}^2} +
\frac{\mu^2}{M_{N_1}}\frac{M_{N_2}}{M_{N_1}}\frac{(M_{N_1}^2-m_{D_1}^2)^2}{(M_{N_1}^2+m_{D_1}^2)^2}+
{\cal{O}}(\mu^3)\,.
\label{mnu2}
\end{equation}
 Eq.~(\ref{mnu1}) shows that the neutrino mass is suppressed by an extra factor $\mu/M_{N_1}$, so that the smallness of neutrino masses, and the 
argument of no fine tuning, do not require tiny Yukawa couplings. 
As for the first term in Eq.~(\ref{mnu2}), it has the standard neutrino 
mass form, i.e. with 2 Dirac masses
in the numerator and one Majorana mass in the denominator, but unlike the usual Seesaw formula, 
it involves only the product of 2 different Dirac masses. Therefore,  
if one of them is smaller than the other, e.g. $\mu << m_{D_1}$, a 
small neutrino mass can be obtained here too with a large Yukawa 
coupling in $m_{D1}$, and no fine-tuning.
As for the second term in Eq.~(\ref{mnu2}), which involves the independent 
parameter $M_{N_2}$,  it also leads to suppressed neutrino masses, even if $M_{N_2}$ largely breaks lepton number.
Now, in the limit $\mu \rightarrow 0$ the point is that the coefficient of the $d=5$  operator vanishes but that of the $d=6$ operator does not.
This can be seen from the fact that the $d=6$ operator takes the form $(Y_N)^\dagger (M_N^{-2}) (Y_N)$, see above, and
doesn't vanish in this limit.
Eq.~(\ref{cd6}) in all cases above,
with for example $m_{D_1}=Y_1v \sim v$ and $M_{N_1}\sim 1$ TeV, 
becomes 
simply $|Y_1|^2/M_{N_1}^2 \sim 1/M_{N_1}^2$ which is large.
The one left-handed plus two right-handed neutrino example above can be 
generalized to the 3 left-handed plus 3 right-handed neutrino above.
The condition for having vanishing neutrino masses is to start with a 
6 by 6 mass matrix which has rank 3. Assuming that all entries of the Yukawa coupling matrix are independent (i.e. barring cancellations between the various entries), it turns out that there is only one possibility to have large Yukawa couplings with three massless light neutrinos and three massive right-handed neutrinos. In the basis $(\nu_e,\nu_\mu,\nu_\tau,N_1,N_2,N_3)$ it is
\begin{equation}
\left( \begin{array}{cccccc}0 & 0 & 0 & c & 0 & 0\\ 
0 & 0 & 0 & d & 0 & 0 \\
0 & 0 & 0 & e & 0 & 0 \\
c & d & e & f & g & a \\
0 & 0 & 0 & g &b & 0 \\
0 & 0 & 0 & a & 0 & 0
\end{array} \right),
\label{matr33}
\end{equation}
plus permutations. This matrix has the particularity that only one of the 3 right-handed neutrinos couples
to light neutrinos at leading order (just as the 1 $\nu$ plus 2 $N$ case above).
From a simple lepton number assignment there is only one way to justify this pattern, which gives in addition
$f=g=0$, i.e. by taking $L_{\nu_e}=L_{\nu_\mu}=L_{\nu_\tau}=L_{N_1}=-L_{N_3}=1$ and $L_{N_2}=0$\,\footnote{For completeness, it can be noted that the 3 light $\nu$s plus 2 heavy $N$ case also leads to a unique possible texture. It corresponds to take no $N_3$, i.e. $a=0$, and requires to take $b=0$ in addition. It can be justified from a L assignment if moreover $f=0$ with $L=1$ for all particles except $N_2$ which has $L=-1$.}$^,$\,\footnote{ During the completion of this work, Ref.~\cite{smirnov} appeared, which also considers this particular texture.}.
The matrix of Eq.~(\ref{matr33}) can be perturbed in many ways:
 \begin{equation}
\left( \begin{array}{cccccc}0 & 0 & 0 & c & \varepsilon_1 & \varepsilon_2\\ 
0 & 0 & 0 & d & \varepsilon_3 & \varepsilon_4 \\
0 & 0 & 0 & e & \varepsilon_5 & \varepsilon_6 \\
c & d & e & f & g & a \\
\varepsilon_1 & \varepsilon_3 & \varepsilon_5 & g &b & \varepsilon_7 \\
\varepsilon_2 & \varepsilon_4 & \varepsilon_6 & a & \varepsilon_7 & \varepsilon_8
\end{array} \right),
\end{equation}
To have two massive light neutrinos, at least one $\varepsilon_i$ among $\varepsilon_{1,...,7}$ must be different from 0.
To have 3 massive light neutrinos, at least two well chosen $\varepsilon_i$ must be different from 0, for 
example $\varepsilon_3$ and $\varepsilon_6$. It is beyond the scope of the present analysis 
to determine all possible perturbations textures which may accommodate the neutrino data along these lines, see also Ref.~\cite{smirnov}. There are many possibilities, and the point is that all of them do lead to unsuppressed $d=6$  operators (i.e. with non-vanishing coefficients in the limit in which all $\varepsilon_i=0$) as long as $a$ and $b$, together with at least one parameter among $c$, $d$, $e$, are different from 0.

\end{document}